\documentclass[journal]{IEEEtran}
\usepackage{graphicx}
\usepackage{amsmath}
\usepackage{algorithm}
\usepackage{hyperref}
\usepackage{booktabs} % For formal tables
\usepackage{tabularx} % For adjustable-width columns

\usepackage{booktabs} % 使用三线表
\usepackage{tabularx} % 允许列宽自动调整
\usepackage{caption}  % 更好的标题控制

\usepackage{array}
\usepackage{makecell}
\usepackage{textgreek}
\usepackage{comment}
\usepackage{bm}
\usepackage{booktabs}
\usepackage{color}
\usepackage{tabularx}
\usepackage{algorithmicx}
\usepackage{algpseudocode}
\usepackage{subcaption}
\algnewcommand\Input{\item[\textbf{Input:}]}  % Define Input
\algnewcommand\Output{\item[\textbf{Output:}]}  % Define Output

\hypersetup{
    colorlinks=true, % 设置链接颜色而不是边框颜色
    linkcolor=blue, % 内部链接颜色设置为蓝色
    filecolor=magenta, % 文件链接颜色设置为洋红色
    urlcolor=cyan, % URL链接颜色设置为青色
    citecolor=blue % 引用链接颜色设置为蓝色
}
\begin{document}
\title{Delay-Effective Task Offloading Technology in Internet of Vehicles: From the Perspective of the Vehicle Platooning}
\author{Kan Yu, \emph{Member, IEEE}, Fuze Zhu, Xiaowu Liu, \emph{Member, IEEE}, Zhiyong Feng, \emph{Senior Member, IEEE}, and Dong Li, \emph{Senior Member, IEEE}
        % <-this % stops a space
\thanks{This work is supported by the National Natural Science Foundation of China with Grant 62301076, the Macao Young Scholars Program with Grants AM2023015, and the Science and Technology Development Fund, Macau SAR, under Grant 0029/2021/AGJ.}
\IEEEcompsocitemizethanks{\IEEEcompsocthanksitem
K. Yu is with the School of Computer Science and Engineering, Macau University of Science and Technology, Taipa, Macau, 999078, P. R. China; the Key Laboratory of Universal Wireless Communications, Ministry of Education, Beijing University of Posts and Telecommunications, Beijing, 100876, P.R. China. E-mail: kanyu1108@126.com. \protect\\
F. Zhu is with School of Computer Science, Qufu Normal University, Rizhao, P.R. China. E-mail: zhufuze1025@163.com\protect\\
X. Liu (the corresponding author) is with School of Computer Science, Qufu Normal University, Rizhao, P.R. China. E-mail: liuxw@qfnu.edu.cn.\protect\\
%D. Chen (the corresponding author) is with the Education, Tianjin Ren'ai College, Tianjin, 301636, P.R.China. E-mail: chendianxia@tju.edu.cn.\protect\\
Z. Feng is with the Key Laboratory of Universal Wireless Communications, Ministry of Education, Beijing University of Posts and Telecommunications, Beijing, 100876, P.R. China. E-mail: fengzy@bupt.edu.cn.\protect\\
D. Li is with the School of Computer Science and Engineering, Macau University of Science and Technology, Taipa, Macau, 999078, P. R. China. E-mail: dli@must.edu.mo.\protect\\

}}

\markboth{IEEE Transactions on Wireless Communications,~Vol.~, No.~, 2024}%
{Shell \Baogui Huang{\textit{et al.}}: Shortest Link Scheduling Under SINR}

\maketitle
\begin{abstract}
The task offloading technology plays a crucial vital role in the Internet of Vehicle (IoV) with the demands of delay minimum, by jointly optimizing the heterogeneous computing resources supported by the vehicles, roadside units (RSUs), and macro base stations (MBSs). In previous works, on the one hand, they ignored the wireless interference among the exchange and sharing of the task data. On the other hand, the available resources supported by the vehicles that have similar driving behaviors, which can form a vehicle platooning (VEH-PLA) and effectively integrate the resources of individual vehicle, has not been addressed. In addition, as a novel resource management paradigm, the VEH-PLA should consider the task categorization, since vehicles in VEH-PLA may have the same task offloading requests, which also  has not attracted enough attention. In this paper, considering the wireless interference, mobility, VEH-PLA, and task categorization, we propose four kinds of task offloading models for the purpose of the processing delay minimum. Furthermore, by utilizing centralized training and decentralized execution (CTDE) based on multi-agent deep reinforcement learning (MADRL), we present a task offloading decision-making method to find the global optimal offloading decision, resulting in a significant enhancement in the load balancing of resources and processing delay. Finally, the simulations demonstrate that the proposed method significantly outperforms traditional task offloading methods in terms of the processing delay minimum while keeping the resource load balancing.
\iffalse
In the dynamic and densely connected domain of the Internet of Vehicles (IoV), efficient management of heterogeneous resources and task offloading is crucial. Different from previous works, on one hand, we consider the types of tasks offloaded by vehicles and classify them according to different task types. On the other hand, we propose task offloading decision algorithms that can adapt to different task types. Specifically, considering the mobility of vehicles, task categorization, and wireless interference, we propose a unique task offloading scheme. First, considering the mobility of vehicles and wireless signal interference in vehicular networks, a truncated Gaussian distribution is used to simulate more realistic vehicle speeds, and by considering unexpected signals and path loss, a more accurate interference model is established, leading to the development of four task offloading models. Subsequently, we analyzed the differences in task types generated in different environments and classified tasks accordingly to reduce data redundancy. Moreover, we proposed a task offloading decision method called CTDE-MADRL-HCRA. This method utilizes centralized training and decentralized execution (CTDE) based multi-agent deep reinforcement learning (MADRL) strategy to optimize task offloading decisions in heterogeneous edge computing environments. Finally, through simulation experiments, we demonstrate that the proposed CTDE-MADRL-HCRA method significantly outperforms traditional methods in terms of processing delay, while maintaining resource load balancing.
\fi
\end{abstract}
\IEEEpeerreviewmaketitle

\section{Introduction}
\label{sec:Introduction}
In the era of intelligent transportation systems, the Internet of Vehicles (IoV) emerges as a transformative technology paradigm, reshaping the way we perceive and interact with transportation infrastructure and vehicles \cite{9759989,9173570}, 
followed by large amounts of computing tasks that requires significant computational resources to process this data in real-time.
However, the limited computing resources and energy constraints of vehicles still pose challenges for executing complex task offloading, which refers to the local task processing or the process of transferring computational tasks from vehicles to nearby edge servers and cloud infrastructure. Accordingly, an effective method of task offloading helps alleviate the computational burden on vehicles by leveraging the computing resources available in the network infrastructure.

In the context of task offloading, the most two crucial aspect lie in the offloading decision and resource management. On the one hand, tasks should be offloaded to a suitable offloading destination, based on factors such as task characteristics, network conditions, performance requirement, and resource availability. On the other hand, 
computational resource allocation in edge servers or cloud data centers should accommodate task offloading requests and ensure timely task execution with minimal resource contention. Currently, heuristic algorithms \cite{9043503}, game-theoretic \cite{10061628}, and methods in \cite{10225421} are popular ones to solve the issue. But these methods failed to work well in a dynamic environment, since they rely on pre-defined rules and complex computations that may not reflect real-time data accurately. 
The deep reinforcement learning (DRL) can continuously optimize offloading schemes and achieve the global performance optimization. In particular, centralized training and decentralized execution (CTDE) is a typical framework in DRL and utilizes the strengths of both by training a global model centrally while allowing agents to execute decisions locally, and corresponding offloading schemes based on DRL and CTDE framework can better address the problem of task offloading in the context of IoV. 

However, due to the size difference of resources and the dynamic nature of IoV environments, above studies on the optimization of the offloading decision and resource management mainly face the following challenges: 1) \emph{resource type}: In fact, vehicles moving on the same road typically exhibit similar behaviors, which can form a platooning mode and be regarded as a novel resource management paradigm; 2) \emph{wireless interference}: The data exchange and sharing among different devices involves wireless communication interference, which further deteriorates the processing delay and success probability of the task offloading; 3) \emph{task categorization}: The vehicles moving on the same road potentially have similar task offloading requirements, the public features of task offloading should be considered. In this way, combining with the concept of VEH-PLA, the resource utilization can be further enhanced.

In view of the above three limitations on the current
works and three challenges for the problem of task offloading, combining with the concept of VEH-PLA, we provide a CTDE-based DRL task offloading decision by optimizing the computational resources to minimize the processing delay and keep the load balancing of the resources, and make a comparison with the
popular offloading decision schemes. The main contributions
of the paper are summarized as follows.
\begin{itemize}
    \item For the problem of task offloading in an context of the IoV, we establish an innovative heterogeneous resource management paradigm with the concept of VEH-PLA. Based on above paradigm, both the minimum processing delay and load balancing of resources are achieved by optimizing computational resources with proposed four kinds of task offloading schemes, under constraints on the vehicular mobility, task categorization, and  interference.
    \item The established resource management paradigm can be modeled as a decentralized partially observable markov decision process (Dec-POMDP). Furthermore, we propose an effective task offloading method by using MADRL based on CTDE, called as a heterogeneous computational resources allocation (CTDE-MADRL-HCRA), to find the global optimal offloading decision, which  integrates the advantages of VEH-PLA, dynamic learning capability of MADRL, leading to a significant performance enhancement in the accuracy of offloading decisions, resource allocation and processing delay.
    \item Simulation results demonstrate the effectiveness and correctness of the proposed schemes. By making a performance comparison between the schemes based on value decomposition networks (VDN), independent Q-learning (IQL) and Q-Mixing networks (QMIX) with the consideration of VEH-PLA and task categorization or not, the CTDE-MADRL-HCRA  shows a superior performance on the processing delay minimum and load balancing of resources, and  provides an effective solution for task offloading in complex scenarios caused by the vehicular mobility and wireless interference.
\end{itemize}

The remainder of this paper is organized as follows. Section \ref{sec:Related Work} summarizes the related work on vehicular network task offloading conducted domestically and internationally. Section \ref{sec:SYSTEM MODEL AND PROBLEM STATEMENT} describes the system model and formulates the VEC task offloading optimization problem. The problem is transformed into a Dec-POMDP, and a learning algorithm based on CTDE is proposed in Section \ref{sec:Distributed Computation Offloading Algorithm Based on CTDE}. Comprehensive simulation results demonstrating the performance are presented in Section \ref{sec:Simulation Results}. Finally, Section \ref{sec:CONCLUSION AND FUTURE WORK} concludes the paper.

\section{Related Works}
\label{sec:Related Work}
The effective computational resource allocation affects the processing delay and success probability of a task offloading. Usually, the computational resources can be supported by the edge computation power (such as the MBSs, RSUs and vehicles) and even cloud center.
Due to the mobility of the vehicles, huge amounts of task offloading, and limited size of computational resources in an IoV, on the one hand, it is challenging to ensure the processing delay and success probability of task offloading. 
%On the other hand, to improve the efficiency of utilization of computational resources and strike a balance between the processing delay and the number of computing tasks satisfied successfully, artificial intelligence (AI) technology can enable the task offloading decision in the IoV by adopting continuous iterative learning, such as the reinforcement learning (RL) \cite{9916276,10175025}, DRL \cite{9047880,9740530}, and federated learning \cite{10324375,9795872}. 
Subsequently, in terms of the computational resources provided by different edge servers, task offloading decision methods, performance metrics of system cost, processing delay, and energy consumption, we survey the methods for task offloading decision in the IoV.

\subsection{Computational Resources Supported by RSUs/MBSs}

%xxxxx. offloading methods (or AI), performance: processing delay, load balancing, energy consumption etc., conclusions
The mobile edge computing (MEC) functions can be realized by integrating edge servers of RSUs/MBSs. In this way, the data processing and storage can be performed directly on these servers and the processing delays can be reduced. Subsequently, in \cite{9047880}, Luo \emph{et al.} deployed RSUs based edge computing along roads to provide enough bandwidth for computing task offloading requiring by vehicles, for the purpose of minimizing the data processing costs while ensuring delay constraints. In particular, they modeled the data scheduling as a DRL problem, which can be solved by an improved deep Q-network (DQN) algorithm. 
By integrating the resources of MBS and RSU, 
In \cite{Sonmez}, Sonmez \emph{et al.} proposed a vehicle edge orchestrator based on two-stage machine learning, under the constraints of the task offloading success probability and task processing delay. Similar works were also done in \cite{cui2023multi}.

Due to the high-speed mobility of the vehicles, the effective communication duration between the vehicles and MBSs/RSUs cannot be enough to support achieve task offloading, since the corresponding processing delay may be higher than the effective communication duration.

\subsection{Computational Resources Supported by vehicles}
Achieving task offloading by exploiting the computational resources of multiple vehicles is effective. For example, by offloading the task to the vehicles moving along the same direction of the road, it is possible to achieve task offloading while ensuring a longer effective communication duration. In \cite{8436044}, Qiao \emph{et al.} regarded the moving vehicles as mobile edge computing resources with the purpose of establishing a collaborative distributed computing architecture, under which a collaborative task offloading and output transmission mechanism was proposed to ensure the processing delay and application-level performance. 
In \cite{9181432}, Xiong \emph{et al.} proposed a heterogeneous intelligent task offloading framework based on three types of V2X communication technologies. 

Considering the computational resources provided by the vehicle edge computing (VEC) server, local vehicles and other vehicles, 
to enhance the delay benefits and energy cost benefits, in \cite{9552188}, Huang \emph{et al.} proposed a task offloading and resource allocation method by jointly designing of task type and vehicle speed, under which the utility function of the vehicle can be derived based on the energy costs and task execution revenue. Finally, a joint optimization scheme for task offloading and resource allocation was developed to maximize the utility level of vehicles constrained by the processing delay, computing resources, and wireless resources. Similar works were also done in \cite{10091540}.

Above works have considered the computational resources supported by the vehicles, and provided an available solution to deal with task offloading interrupt existing in offloading centered on the RSUs. But the coordinated task offloading framework among vehicles and collaboration were ignored.

\subsection{Computational Resources Supported by VEH-PLA}
Different from randomly moving vehicles, the vehicles in VEH-PLA are featured by the same driving behaviors, such as same driving direction, smooth speeds, and tight following distance \cite{xiao}. As an innovative method through V2V communication technology that enables the coordinated operation among multiple vehicles, the VEH-PLA offers a stable topology to provide a larger size of computational resources and effective communication duration \cite{xiaotingting}. In addition, the task offloading by using the vehicular resources can further decrease the corresponding processing delay.

In \cite{xiao}, Xiao \emph{et al.} redistributed complex tasks via VEH-PLA, thereby optimizing the utilization of VEH-PLA's idle resources for enhanced task processing. This approach benefited from the inherent stable connectivity, adjustable maneuverability, and sufficient charging capabilities characteristic of VEH-PLA. Moreover, they introduce a Stackelberg game model to describe the interplay between mobile edge platoon cloud (MEPC) platforms and vehicles requiring task offloading, focusing on maximizing their combined utility. 
In \cite{10130367}, Zheng \emph{et al.} analyzed the processing delay and energy consumption of local offloading and VEH-PLA offloading when offloading the task to VEH-PLA members, and proposed a method to minimize the long-term maximum task queue by optimizing the allocation of communication and computing resources. Finally, the effectiveness of the proposed method was demonstrated through online algorithms using Lyapunov optimization and successive convex approximation (SCA) methods.
In \cite{9266767}, Chao \emph{et al.} designed a contract based incentive mechanism to incentivize vehicles to form VEH-PLA and share their computing resources. Then, for maximizing vehicle utility, a joint optimization of offloading decisions, communication, and computing resource allocation schemes of the vehicle and BS was formulated.
The problem of task offloading failure caused by potential disconnections among vehicles was studied in \cite{maxiaoting}, a multi-leader multi-follower Stackelberg game was introduced for the co-optimization of pricing strategies and computational resource distribution, with the goal of maximizing participant revenues. Similar works were also done in \cite{9951056}.

However, for one thing, all above studies ignored wireless interference over the processing of task offloading, which affects the QoS of communication channel and the processing delay,
resulting a failure of the task offloading. For another thing, facing a large number of task offloading requests, there has been no analysis of task types, particularly the differences between tasks managed in a queue environment and those performed by traditional road vehicles. VEH-PLA, as a new type of computational resource allocation framework, allows its members to drive collaboratively, collectively completing public tasks such as environmental perception, decision-making, and emergency response. This can effectively reduce the frequency of task offloading and alleviate the pressure of insufficient computing resources in the IoV.

\section{SYSTEM MODEL AND PRELIMINARIES}
\label{sec:SYSTEM MODEL AND PROBLEM STATEMENT}
\iffalse
In this section, we introduce a novel heterogeneous edge computing framework of a IoV based on the concept of VEH-PLA. Then, we establish a vehicular moving model by introducing a more realistic truncated Gaussian distribution, and performance metrics of the task offloading. Finally, the optimization problem of task offloading enabled by VEH-PLA is formulated, with the purpose of minimizing the processing delay of the computing task while ensuring the load balancing.  
\fi
\iffalse
\begin{figure}[htbp]
  \centering % 让图片居中显示
  \includegraphics[width=0.5\textwidth]{无任务卸载.pdf} % 图片文件的路径
  \caption{Vehicular Network Road Scenario} % 图片的标题
  \label{fig:network model} % 用于引用图片的标签
\end{figure}
\fi

\subsection{System Model}

%To improve driving safety and traffic efficiency, moving vehicles need the data processing，task, task offloading, types of tasks: public task and personalized task, offloading methods/available resources: local, other vehicels, RSU, MBS, as shown in Fig. 1, 

\subsubsection{Computational resource architecture}
As shown in Fig. \ref{fig:network model}, the management architecture of computational resources in the context of the IoV includes a set of MBSs, denoted by $S_{mbs}=\{mb_1, mb_2,..,mb_{n_{mbs}}\}$, a set of RSUs, denoted by $S_{rsu}=\{rs_1, rs_2,..,rs_{n_{rsu}}\}$
and a set of moving vehicles, denoted by $S_{ve}=\{ve_1, ve_2,..,ve_{n_{ve}}\}$, where $n_{mbs}$, $n_{rsu}$, $n_{ve}$ denote the number of MBSs, RSUs and vehicles, respectively. The MBSs are distributed randomly at different locations keeping away from the road, while the RSUs are placed equidistantly along the roadside. In addition, the vehicles, RSUs, and MBSs poss limited computational resources. Without losing generality, let $F^{V}_{\max}$, $F^{R}_{\max}$, and $F^{M}_{\max}$ denote the maximum computing capabilities of individual vehicle, RSU, and MBS (\emph{i.e.}, the maximum CPU cycle frequency), respectively.

\begin{figure}[htbp]
  \centering % 让图片居中显示 
  \includegraphics[width=0.35\textwidth]{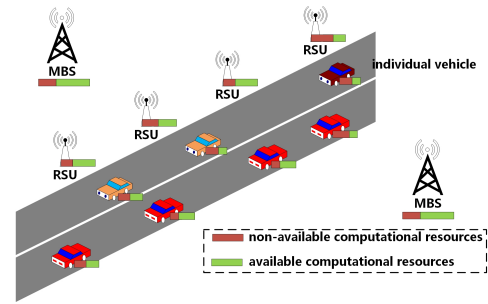} % 图片文件的路径
  \caption{Resource management architecture} % 图片的标题
  \label{fig:network model} % 用于引用图片的标签
\end{figure}

\subsubsection{Task offloading}
To improve the driving safety and traffic efficiency, moving vehicles generate the tasks needed being computed by occupying computational resources. Generally speaking, the generated tasks usually are classified into public tasks, such public demands of traffic sign recognition and road obstacle perception, as well as personalized tasks for satisfying individual differentiated demands. Because of limited computational resources and affordable processing delay of individual vehicle, vehicles requesting for a task can resort to local computational resources or available ones of other vehicles, RSUs as well as MBSs, as shown in Fig. \ref{fig:network model}. 

\subsubsection{VEH-PLA based resource management}
\begin{figure}[htbp]
  \centering % 让图片居中显示 
  \includegraphics[width=0.35\textwidth]{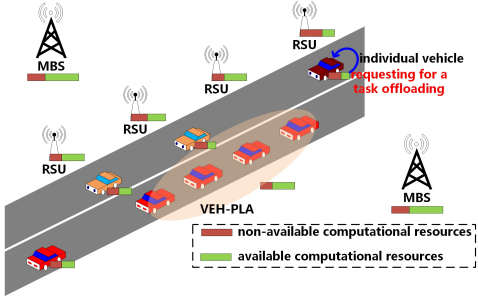} % 图片文件的路径
  \caption{VEH-PLA based resource management architecture} % 图片的标题
  \label{fig:2} % 用于引用图片的标签
\end{figure}

It is worth mentioning that, considering the available computational resources provided by other vehicles, VEH-PLA is a significant resource management architecture to effectively integrate the resources of the individual vehicle. Constrained by traffic regulations (such as driving direction, safe distance, speed limits), vehicles driving on the same road typically exhibit similar behaviors. Based on this, these vehicles moving a local range can form a platooning mode, driving together in an orderly manner according to certain rules and distances (especially on highways during peak hours or long-distance journeys). In addition, the VEH-PLA can expand its size through lane changing and merging, achieving more efficient communication and resource coordination optimization. Let $\bar{S}_{ve}=\{ve_1, ve_2,..,ve_{m_{ve}}\}$ be the set of vehicles in the same VEH-PLA, where $m_{ve}$ ($<n_{ve}$) is the number of vehicles in the platooning.
In this way, considering task types and VEH-PLA, feasible task offloading decisions can be provided by the individual vehicle, VEH-PLA, RSUs and MBSs, as shown in Fig. \ref{fig:2}.

In addition, for better describing task offloading, we discretize the time duration $T$ into $N$ decision episodes, indexed by $t$, with the length of $T/N$. It is assumed that the vehicles generate a task at most at the episode $1\leq t\leq T/N$.

\subsection{Vehicle Speed Model}
Without loss of generality, we employ a Cartesian coordinate system and define the positive direction as moving towards the right of the Fig. \ref{fig:network model}, with coordinates measured in meters. The position of $i$-th vehicle at time slot $t$ is $\textbf{z}^{i}_{ve}(t)=[x_{ve_i}, y_{ve_i}]^\textbf{T}$. Similarly, the locations of $j$-th RSU and $k$-MBS can be denoted by $\textbf{z}^{j}_{rsu}(t)=[x_{rsu_j}, y_{rsu_i}]^\textbf{T}$ and $\textbf{z}^{k}_{mbs}(t)=[x_{mbs_k}, y_{mbs_k}]^\textbf{T}$, respectively.

Using a truncated Gaussian distribution model to simulate vehicle traffic is more in line with the actual movement characteristics of vehicles \cite{8745530}, where the mean is $\mu$ and the standard deviation is $\sigma$. Let $v_{\text{min}} = \mu - 3\sigma$ and $v_{\text{max}} = \mu + 3\sigma$ be the minimum and maximum truncation ranges for speed, and let the speed be represented by $v_i$. To ensure the consistency of speed among members of VEH-PLA, the generated speeds need to be synchronized. In this way, the speed of member-vehicles, denoted by $v_{\bar{S}_{ve}}$, is set to the average speed of the whole platooning, namely $v_{\bar{S}_{ve}}=\frac{1}{m_{\rm ve}}\sum^{m_{\rm ve}}_{i=1}v_{i}$.
\iffalse
Define the coverage area of the RSU as a disc with the maximum radius of $L^{\max}_{rsu}$. The effective communication duration between $j$-th RSU and $i$-th moving vehicle, denoted by $t^{r_j}_{ve_i}$, can be calculated by
\begin{equation}
t_{ve_i}^{r_j} = 
\begin{cases} 
\frac{x_{rsu}^j - x_{ve}^i + \sqrt{L_{rsu}^{\max} - y_{rsu}^j}}{v_i}, & \text{if } v_i > 0 \\[10pt]
\frac{x_{rsu}^j - x_{ve}^i - \sqrt{L_{rsu}^{\max} - y_{rsu}^j}}{v_i}, & \text{if } v_i < 0
\end{cases}
\end{equation}
where $v_i > 0$ indicates that vehicle $\text{ve}_i$ is moving in the positive direction along the road, while $v_i < 0$ signifies that vehicle $\text{ve}_i$ is traveling in reverse along the road.
Similarly, let \( L_{ve}^{\max} \) be the maximum communication distance among the vehicles. Considering the moving direction, relative velocities $(v_i, v_j)$, and positions $(z_{ve}^i, z_{ve}^j)$ of the vehicles,  the effective communication duration between $i$-th vehicle and $j$-th vehicle, denoted by $t_{{ve}_i}^{{ve}_j}$ can be calculated as 
\begin{equation}
t_{ve_i}^{ve_j} = 
\begin{cases} 
\frac{x_{{ve}_j} - x_{{ve}_i} + L_{ve}^{\max}}{v_i - v_j}, & \text{if } v_i > v_j \text{ (same direction)} \\
\frac{x_{{ve}_i} - x_{{ve}_j} + L_{ve}^{\max}}{v_i - v_j}, & \text{if } v_i < v_j \text{ (same direction)} \\
\frac{x_{{ve}_j} - x_{{ve}_i} + L_{ve}^{\max}}{v_j + v_i}, & \text{otherwise}
\end{cases}
\label{eq:3}
\end{equation}
\fi

\subsection{Advantages of VEH-PLA and Task Offloading Methods}\label{sub:task offloading methods}
\begin{figure}[htbp]
  \centering % 让图片居中显示 
  \includegraphics[width=0.35\textwidth]{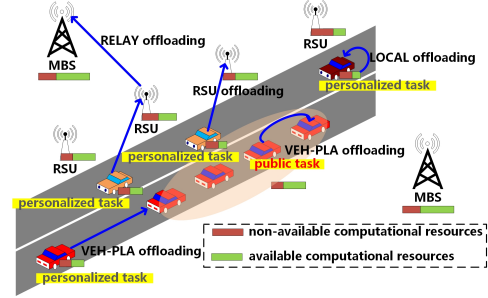} % 图片文件的路径
  \caption{Four kinds of task offloading models} % 图片的标题
  \label{fig:task offloading models} % 用于引用图片的标签
\end{figure}

Vehicles driving on the same road usually have similar task processing requirements (e.g., the road condition information, road congestion information, and weather information etc). 
The framework VEH-PLA has a stable topology structure and tight following distance. Accordingly, on the one hand, tasks requesting from vehicles in the platooning can be regarded as public tasks, and can be handled uniformly by using the same resources. While the individual differentiated task requests can be called as personalized tasks, which need allocate additional resources to process the tasks. 
On the other hand, the VEH-PLA framework also extends the available computing resources of individual vehicle, increasing the success probability of task offloading. Based on the advantages of VEH-PLA, when the vehicle requests for a task offloading,  based on the constraints of the task types, size of available computing resources and processing delay provided by different offloading models, the vehicle will select the best offloading decision among the \textbf{LOCAL offloading}, \textbf{VEH-PLA offloading}, \textbf{RSU offloading}, and \textbf{RELAY offloading}, as shown in Fig. \ref{fig:task offloading models}.

In detail, the \textbf{VEH-PLA offloading model} can support two kinds of task types, i.e., public tasks and personalized tasks. On the one hand, the vehicle requesting for a task offloading belongs to the platooning, it first determines whether the task is public or not. If the task is public, it does not occupy any resources to compute the task and directly gets the result from the lead vehicle of the VEH-PLA. Otherwise, it will perform task offloading. On the other hand, when the $i$-th vehicle does not belong to the platooning, if effective communication duration between $ve_i$ and any one of vehicles in the platooning is less than or equal to the affordable offloading delay of $i$-th vehicle, and the latter has enough computing resources satisfying the demand of $i$-th vehicle's task, then $ve_i$ can execute the \textbf{VEH-PLA offloading model}.
\iffalse
In terms of the minimum processing delay provided by different offloading models, if the available computing resources of LOCAL and VEH-PLA offloading models are not sufficient or cannot provide the resources required by $i$-th vehicle's computing task, the vehicle may consider the \textbf{RSU offloading model}, based on constraints on computing resources and processing delay. Otherwise, if above three offloading models cannot achieve an expected processing delay, the RSU can be regarded as the relaying node, and transfer the computing task of the $i$-th vehicle to one of available MBSs, that is, executing \textbf{MBS offloading model}.
\fi
\subsection{The Delay Analysis of the Task Offloading Model}
As a result of these diverse offloading models, differentiated demand of vehicles, different scale of computing resources, the processing delay of a task offloading consists of different types of delays for LOCAL, VEH-PLA, RSU and RELAY offloading models.

\iffalse
Specifically, based on the concept of task offloading methods, the processing delay of \textbf{LOCAL} only includes the offloading delay, while the processing delays of \textbf{VEH-PLA} and \textbf{RSU} consist of both the communication delay and offloading delay, and that of \textbf{RELAY} includes the  communication delay, offloading delay and relaying delay.
When the computing tasks of vehicles are offloaded to themselves, vehicles in the platooning, RSUs and MBSs, the transmission rate supported by vehicles requesting for task offloading serves as a crucial component of affecting the processing delay. The channel conditions, communication distance, and transmission power, particularly the cumulative interference generated by non-intended communications are key factors impacting the transmission rate. Consequently, the accurate construction of a transmission rate and offloading model, considering wireless conditions and interference, is of significant importance for optimizing resource managements and satisfying differentiated demand in the IoV system. 
\fi

\subsubsection{Transmission Rate between Vehicles and RSUs} 
Given that $i$-th RSU $rs_i$ can keep effective communication duration with $k$-th moving vehicle $ve_k$. In other words, $k$-th vehicle can offload the computing task to $i$-th RSU before moving out the latter's coverage area. Combining with the conclusions in \cite{10155623,9487018}, the transmission rate between $ve_k$ and $rs_i$, denoted by $R_{ve_k,rs_i}$, can be represented as
\begin{equation}
R_{ve_k,r_i}=W\log_2(1+\frac{p_{ve_k}h_{ve_k,rs_i}d_{ve_k,rs_i}^{-\alpha}}{\sigma^{2}_{rs_i}+\sum\limits^{n_{ve}}_{j=1,j\neq k}\frac{p_{ve_j}h_{ve_j,rs_i}}{d_{ve_j,rs_i}^{\alpha}}\cdot\delta_{ve_j,rs_i}})
\label{eq.IR}
\end{equation}
where $W$ denotes the transmission bandwidth of the vehicles, $p_{ve_k}$ is the transmission power of $k$-th vehicle, $h_{ve_k,rs_i}$ and $d_{ve_k,rs_i}$ are the channel fading coefficient and the distance between $k$-th vehicle and $i$-th RSU. In particular, $h_{ve_k,rs_i}$ follows an exponential distribution with unit mean.
$\alpha$ is the path-loss exponent, and $\sigma^{2}_{rs_i}$ is the additive Gaussian white noise (AGWN) power at the $i$-th RSU. 
Moreover, the indicator function $\delta_{ve_j,rs_i} \in \{0,1\}$ represents whether the $j$-th vehicle is associated with $i$-th RSU or not. Specifically, $\delta_{ve_j,rs_i}=1$, if the $j$-th vehicle requests a task offloading and is within the coverage of the $i$-th RSU; otherwise, $\delta_{ve_j,rs_i}=0$, namely the $j$-th vehicle does not be with no need for a task offloading, or is not within the coverage range of $i$-th RSU.

\subsubsection{Transmission Rate among Vehicles}
Given that $k$-th vehicle (in the VEH-PLA or not) offloads its task to an available vehicle, denoted by $i$-th one, in the VEH-PLA.
That is, the latter can provide enough computing resources to the former one. 
Transmission rate between the $k$-th vehicle and the $i$-th vehicle, denoted by $R_{ve_k,ve_i}$, can be represented as
\begin{equation}
R_{ve_k,ve_i}=W\log_2(1+\frac{p_{ve_k}h_{ve_k,ve_i}d_{ve_k,ve_i}^{-\alpha}}{\sigma^2_{ve_i}+\sum\limits^{n_{ve}}_{j=1,j\neq 
k,j\neq i}\frac{p_{ve_j}h_{ve_j,ve_i}}{d_{ve_j,ve_i}^{\alpha}}\cdot\delta_{ve_j,ve_k}})
\label{eq:transmission rate beween vehicles}
\end{equation}
where $\sigma^2_{ve_i}$ denotes the AWGN power at the $i$-th vehicle, and the indicator function $\delta_{ve_j,ve_k} \in \{0,1\}$ denotes whether the $j$-th vehicle and the $k$-th vehicle are both associated with the $i$-th vehicle. Specifically, $\delta_{ve_j,ve_k}=1$, if the $j$-th vehicle sends a task offloading to the $i$-th vehicle, which has been assigned its parts of resources to the $k$-th vehicle for achieving the task data processing; otherwise, $\delta_{ve_j,ve_k}=0$.

\iffalse
In addition, the transmission rates under RELAY and LOCAL  offloading models can be calculated by using similar methods of Eq. \eqref{eq.IR} and Eq. \eqref{eq:transmission rate beween vehicles}, respectively.
\fi

\subsubsection{The Processing Delay Analysis} 
Given the scenario of binary task offloading, based on the constraints of available computing resources and offloading delay, the vehicles with task offloading demands have the option to offload the task locally, to RSUs, to platoon member vehicles or to other RSUs or MBSs by using a RSU as the relay. Due to the mobility characteristics of the vehicles, the supported transmission rate for a given vehicle with the demands of task offloading, denoted as the $k$-th vehicle, are dynamic, resulting in selecting different offloading schemes. Let $\Psi_{ve_k}(t)$ represent the set of candidate offloading decisions for the $k$-th vehicle at time slot $t$, including decisions generated by LOCAL, VEH-PLA, RSU, and MBS offloading models. Consequently,
\begin{equation}
\begin{aligned}
\Psi_{ve_k}(t) &=\underbrace{\{\Psi_{ve_k}^{ve_1}(t),..., \Psi_{ve_k}^{ve_{k-1}}(t),\Psi_{ve_k}^{ve_{k+1}}(t)...,\Psi_{ve_k}^{ve_{n_{ve}}}(t)\}}_{\textbf{LOCAL}~~\textrm{and}~~\textbf{VEH-PLA}}\\
&~~~\cup
\underbrace{\{\Psi_{ve_k}^{rs_1}(t),...,\Psi_{ve_k}^{rs_{n_{rs}}}(t)\}}_{\textbf{RSU}}\\
&~~~\cup\underbrace{\{\Psi_{ve_k}^{mb_1}(t),...\Psi_{ve_k}^{mb_{n_{mb}}}(t)\}}_{\textbf{RELAY}}
\end{aligned}
\end{equation}
where
$\Psi_{ve_k}^m$ is a binary random variable and denotes a offloading decision, namely $k$-th vehicle offloads its computing task to $m$-th node with enough resources. In detail, $\Psi_{ve_k}^m=1$ indicates that the task  of the $k$-th vehicle is offloaded to the $m$-th unoccupied resource node; otherwise, $\Psi_{ve_k}^m=0$. In addition, $\Psi_{ve}(t)$ denotes the set of candidate offloading decisions for all vehicles at time slot $t$. Furthermore, we establish the offloading delay models for four offloading models. Specially, the RSUs and the MBSs are linked through fiber optics, so the offloading delay includes sending delay, rather than transmission delay.

The task generated by a vehicle can be modeled as $\mathrm{\Omega}_{{ve}_k}(t) = \{\lambda_{{ve}_k}(t), \rho_{{ve}_k}(t), D^{\max}_{{ve}_k}(t)\}$, where $\lambda_{{ve}_k}(t)$ denotes the size of the computing task, $\rho_{{ve}_k}(t)$ denotes the processing density or the number of CPU cycles required for per unit of data, and $D^{\max}_{{ve}_k}(t)$ represents the affordable maximum delay requirement of the $k$-th vehicle. Next, we present the corresponding framework of delay analyses for offloading models of LOCAL, VEH-PLA, RSU, and RELAY.

\textbf{1) Processing delay in LOCAL offloading model}: Given the $k$-th vehicle with the demands of task offloading during episode $t$, if it has sufficient resources and satisfies the minimum delay requirement, a LOCAL offloading decision is executed. Then, the processing delay for LOCAL at episode $t$, denoted by $D^{ve_k}_{ve_k}(t)$, is calculated as $D_{{\rm ve}_k}^{{\rm ve}_k}(t) = \frac{\lambda_{{\rm ve}_k}(t) \rho_{{\rm ve}_k}(t)}{F^{V}_{\max}}$.

\textbf{2) Processing delay in VEH-PLA Offloading Model:} If the $k$-th vehicle is within the communication range of the vehicle platoon and the communication delay is less than the effective communication duration, then the task can be offloaded to the $j$-th vehicle that belongs to the vehicle platoon, where the communication delay is $D_{{\rm ve}_k,{\rm ve}_j} = \frac{\lambda_{{ve}_k}(t)}{R_{{ve}_k,{ve}_j}}$. Consequently, the processing delay of the vehicle in the platooning by using VEH-PLA offloading model, denoted by $D_{{ve}_k}^{{ve}_j}(t)$, is calculated as $\frac{\lambda_{{ve}_k}(t)\rho_{{ve}_k}(t)}{F^{V}_{\max}} +  D_{{\rm ve}_k,{\rm ve}_j}$. 

\textbf{3) Processing delay in RSU offloading model:} Enabled by dynamic frequency and voltage scaling techniques for RSUs \cite{dongtaipinlv}, allowing for dynamic allocation of CPU frequency for each task. Let $\zeta_{r_i}\left(t\right)$ denote the total computational resource demands of multiple tasks served by the $i$-th RSU $r_i$. Then, the size of  computational resources for $k$-th vehicle allocated by $i$-th RSU, denoted by,  $f_{r_i}^{{ve}_k}\left(t\right)$, can be obtained as $\left(\frac{\lambda_{{ve}_k}\left(t\right)\rho_{{ve}_k}\left(t\right)}{\zeta_{r_i}\left(t\right)}\right)\ast F^{R}_{\max}$. Therefore, the offloading delay is given by $D_{{ve}_k}^{{tr}_i}\left(t\right)=\frac{\lambda_{{ve}_k}\left(t\right)\rho_{{ve}_k}\left(t\right)}{f_{r_i}^{{ve}_k}\left(t\right)}$. Furthermore, from Eq. \eqref{eq.IR}, the communication delay caused by offloading the task to $i$-th RSU is $D_{ve_k, r_i}(t) = \frac{\lambda_{ve_k}(t)}{R_{ve_k, r_i}}$. To sum up, the processing delay to the RSU can be expressed as 
\begin{equation}
D_{{ve}_k}^{r_i}\left(t\right)=D_{{ve}_k}^{{tr}_i}\left(t\right)+D_{ve_{k}, r_{i}}(t).
\end{equation}

Through optical fiber links, the task is transferred from the $i$-th RSU to the $j$-th RSU. Let $R_{r_i,r_j}$ be the transmission rate between the $i$-th RSU and the $j$-th RSU, then the processing delay can be expressed as

\begin{equation}
D_{{ve}_k}^{r_j}(t) = \frac{\lambda_{{ve}_k}(t)}{R_{{ve}_k,r_i}} + \frac{\lambda_{{ve}_k}(t)}{R_{r_i,r_j}} + \frac{\lambda_{{ve}_k}(t)\rho_{{ve}_k}(t)}{f_{r_j}^{{ve}_k}(t)}.
\end{equation}

\textbf{4) Processing delay in RELAY offloading model}: 
The modeling of the computing resources provided by the $j$-th MBS to vehicle users as a Poisson distribution is denoted by $f_{b_j}^{ve}(t) = F^{M}_{\max} - f_{b_j}^{nve}(t)$, where $f_{b_j}^{nve}(t)$ represents the resources used by non-vehicle users. The rest of the content remains consistent with the relay RSU. Therefore, the processing delay of the k-th vehicle when offloading its task to the j-th MBS can be expressed as

\begin{equation}
    D_{{ve}_k}^{b_j}(t) = \frac{\lambda_{{ve}_k}(t)}{R_{{ve}_k,r_i}} + \frac{\lambda_{{ve}_k}(t)}{R_{r_i,b_j}} + \frac{\lambda_{{ve}_k}(t)\rho_{{ve}_k}(t)}{f_{b_j}^{{ve}_k}(t)}.
\end{equation}

In summary, the processing delay can be universally represented as a summation of delays associated with each offloading model, and can be calculated as follows:
\begin{align}
D_{ve_k}^{\textrm{sum}}(t) = &\underbrace{\Psi_{ve_k}^{ve_k} D_{ve_k}^{ve_k}(t)}_{\textbf{LOCAL}} + \underbrace{\Psi_{ve_k}^{ve_j} D_{ve_k}^{ve_j}(t)}_{\textbf{VEH-PLA}} + \underbrace{\Psi_{ve_k}^{r_i} D_{ve_k}^{r_i}(t)}_\textbf{{RSU}} \nonumber \\
&  + \underbrace{\Psi_{ve_k}^{b_j} D_{ve_k}^{b_j}(t)}_{\textbf{RELAY}}
\end{align}

\subsection{Problem Formulation: Delay-effective Task Offloading}
Based on the above description, the purpose of this paper is to optimize the task offloading decisions of all vehicles demanding for a task offloading at episode $t$, denoted by 
$$\Psi(t) = \{\Psi_1(t),..., \Psi_k(t),..., \Psi_{n_{\text{ve}}}(t)\}.$$ 
Formally, the optimization problem of a delay-effective task offloading can be written as follows
\begin{subequations}\label{eq:optimal problem}
\begin{align}
   \text{Objective:}& \min_{\Psi(t)} \quad  \sum_{k=1}^{n_{ve}} D_{\text{ve}_k}^{\text{sum}}(t) \label{objective}\\
    \text{s.t.} \quad & \psi_{ve_k}^m(t) \in \{0, 1\}, \forall k, \psi_{ve_k}^m(t)\in \Psi_{ve_k}(t) \label{con:C1}\\
    & \sum \psi_{ve_k}^m(t) \leq 1, \forall k \label{con:C2}\\
    & D_{\text{ve}_k}^{\text{sum}}(t) \le D^{\max}_{\text{ve}_k}(t), \forall k \label{con:C3}\\
&\sum_{k=1}^{n_{ve}}\mathrm{\Psi}_{{ve}_k}^{r_i}<N, \forall k, \psi_{ve_k}^{r_i}(t)\in \Psi_{ve_k}(t) \label{con:C4}
\end{align}
\end{subequations}
where constraints of Eq. \eqref{con:C1} and Eq. \eqref{con:C2} indicate that the $k$-th vehicle with demand of a task offloading only chooses one of four available offloading models at episode $t$, as described in Subsection \ref{sub:task offloading methods}. Eq \eqref{con:C3} ensures that the processing delay satisfies the maximum affordable delay requirements of $k$-th vehicle, while Eq. \eqref{con:C4} can achieve the load balancing of the resources on the RSUs, by allowing to serve $N$ vehicles simultaneously at most over all episode.

To obtain a satisfactory task offloading decision, a search must be required in the decision space with the size of ${{(n_{\text{ve}} + n_{\text{rsu}} + n_{\text{mbs}})}^{n_{\text{ve}}}}$. Based on this fact, traditional methods, such as \cite{7608300}, typically necessitate adjusting decisions iteratively to achieve the global optimization, but do not work well in rapidly moving vehicular environments. Facing the challenging in acquiring a priori knowledge of the dynamic environment, the DRL has been shown that it allows for learning optimal offloading schemes via continual interaction with the stochastic vehicular environments \cite{wangyun}. 
\iffalse
The centralized DRL methods faced the issues of scalability when dealing with the exponential growth of the action space generated by a large number of vehicles. While distributed DRL encounters above challenges in coordinating complex behaviors among vehicles to achieve the global optimization.
\fi
Therefore, in Section \ref{sec:Distributed Computation Offloading Algorithm Based on CTDE}, based on a CTDE, we present a multi-agent DRL (MADRL) algorithm  to strike a balance between computational complexity and convergence performance to find a long-term optimal solution, satisfying the processing delay and resource load balancing requirements.

\section{Distributed Computation Offloading Algorithm Based on CTDE}
\label{sec:Distributed Computation Offloading Algorithm Based on CTDE}
The optimal task offloading decisions of the vehicles demanding task offloading in Eq. \eqref{eq:optimal problem} depend on the global observation of the state information for all vehicles. However, due to complex behaviors among vehicles caused by their mobility, it is challenging to obtain the global observation of the state formation. Accordingly, in this section, we first model the task offloading decision problem as a Dec-POMDP, then based on MADRL, which are constrained by local state information observation, affordable processing delay and load balancing, we present a CTDE-MADRL based task offloading method in the context of VEH-PLA to determine the optimal offloading decisions for minimizing the processing delay while keeping the load balancing of resource.

\subsection{Distributed Multi-Agent System Design}
First of all, each agent in IoV can collect or observe the local information in its surrounding environment at episode $t$. Subsequently, the agent independently makes an action choice for maximizing the overall expected shared reward. Formally, based on Dec-POMDP, a distributed vehicular task offloading can be represented as a tuple of $\langle I, S, \{O_i(t)\}_{i \in I}, P, \{A_i(t)\}_{i \in I}, \{r_i(t)\}_{i \in I}, \gamma \rangle$ where $I$ is the set of agents, $S$ represents the set of all possible states of the environment, including all states that the system can reach at any episode, $s(t) \in S$ represents the environmental state at episode $t$, $O_i(t)$ is the local observation space possessed by $i$-th agent at episode $t$, $o_i(t) \in O_i(t)$ is the specific instance of the local observation,
$P$ is the state transition function,  $A_i(t)$ is the action spaces available for $i$-th agent at episode $t$, $a_i(t) \in A_i(t)$ represents the action selected by $i$-th agent, $r_i(t)$ is the reward of $i$-th agent at episode $t$, and $\gamma \in [0,1]$ is the discount factor that balances the present and future rewards. Due to limited observations achieved by each agent, it cannot fully observe the environmental state space $S$ and instead use the local observation $o_i(t) \in \{O_i(t)\}_{i\in I}$ as a substitute for the state $s(t) \in S$. 
In addition, the state transition function can be ignored, namely $P$ can be removed from the tuple, since the task offloading decision problem is modeled as a distributed MADRL task. Therefore, the configurations of distributed vehicular task offloading are rewritten as follows.

%At every time step, each agent can only collect or observe limited information about its local environment, characterized through local environmental observations. Based on an understanding of these local observations, each agent independently makes action choices aimed at maximizing the overall expected shared return. Based on Dec-POMDP, the distributed VEC offloading task can be represented by a tuple \( \langle I, S, \{O_i\}_{i \in I}, P, \{A_i\}_{i \in I}, \{r_i\}_{i \in I}, \gamma \rangle \). Here, \( i \in I \) denotes the set of agents, which in this scenario cannot fully observe the environmental state space \( S \) and instead use the local observation \( o_t^i \in O_t^i \) as a substitute for the state function \( s_t \in S \). The model-free MARL task adopted ignores the state transition function \( p(s'|s, a) \in P : S \times A^1 \times A^2 \ldots \times A^I \times S' \rightarrow [0,1] \). \( \{A_i\}_{i \in I} \) is the action space set for agent \( i \). The parameter \( \gamma \in [0,1] \) is the discount factor that balances the present and future rewards.

\textbf{1) Local Observations:} In terms of the location, task size, computing resource demand, processing delay, and relationship with vehicular platooning, the local observation of $i$-th agent at episode $t$ can be represented as
\begin{equation}
o_i(t) = (\Theta_i(t), \lambda_i(t), \lambda_i(t) \rho_i(t), D_{{ve}_i}^{{ve}_i}(t), \delta_{{ve}_i}^{{\textrm{pla}}_{ve_j}}(t))   
\end{equation}
where $\Theta_i(t)$ denotes the location of $i$-th agent at episode $t$, $\lambda_i(t)$ represents the task size, $\lambda_i(t) \rho_i(t)$ signifies the computing resources needed by the task, $D_{{ ve}_i}^{{ ve}_i}(t)$ is the corresponding processing delay required for executing LOCAL offloading, and $\delta_{{ve}_i}^{{\rm pla}_{ve_j}}(t)\in\{0,1\}$ indicates that whether $i$-th agent makes a connection with $j$-th agent belonging to the vehicular platooning. If the $i$-th agent cannot be connected to any agents existing in the VEH-PLA, then $\delta_{{ve}_i}^{{\textrm{pla}}_{ve_j}}(t)=0$.

\textbf{2) Action Space:} The action space of the $i$-th agent consists of potential offloading decisions. In this way, the set of offloading decisions available to the $i$-th agent at episode $t$ can be represented as
\begin{equation}
\begin{aligned}
A_i(t) &= \Psi_{ve_i}(t) \\
&= \underbrace{\{\Psi_{ve_i}^{ve_1}(t),..., \Psi_{ve_i}^{ve_{i-1}}(t), \Psi_{ve_i}^{ve_{i+1}}(t)..., \Psi_{ve_i}^{ve_{n_{ve}}}(t)\}}_{\textbf{LOCAL}~~\text{and}~~\textbf{VEH-PLA}} \\
&\cup \underbrace{\{\Psi_{ve_i}^{rs_1}(t),..., \Psi_{ve_i}^{rs_{n_{rs}}}(t)\}}_{\textbf{RSU}} \\
&\cup \underbrace{\{\Psi_{ve_i}^{mb_1}(t),...\Psi_{ve_i}^{mb_{n_{mb}}}(t)\}}_{\textbf{RELAY}}
\end{aligned}
\end{equation}

\textbf{3) Reward:} To update task offloading decisions to maximize the whole system rewards, all agents share a common team reward. By using similar methods to deal with actions of agents that cannot satisfy the constraints \cite{reward1}, we set a negative penalty for agents to describe the results when violating constraints of Eq. \eqref{con:C1} - Eq. \eqref{con:C4}. Then the reward for the $i$-th agent based on the objective function and constraints to shape the training process is given by
\begin{equation}
r_i(t) = \begin{cases} 
- \sum_{i \in I} D_{ve_i}^{\textrm{sum}}(t) & \text{if constraints are met} \\
pc\cdot(\eta_1 + \eta_2)\text{} & \text{otherwise}
\end{cases} 
\end{equation}
where $pc$ is the penalty coefficient, while $\eta_1$ and $\eta_2$ are penalties corresponding to constraints of Eq. \eqref{con:C3} and Eq. \eqref{con:C4}, respectively. A well-designed penalty structure for the training process is crucial to accelerate the algorithm convergence.

\iffalse
The local observation information and reward information of the agents are unbalancing and associated with the noise. Using above two raw information directly will significantly deteriorates the convergence speed of the agents, which makes it more challenging to achieve the optimal solution and potentially lead to non-convergence of MADRL. Therefore, the normalization of vehicle location, task size, local computation delay, computing resource demands is needed to rescale the local observation information and reward information, thereby enhancing the learning efficiency of the agents and the robustness of the MADRL model.
\fi
Under the framework of the cooperative MADRL, the agents collectively execute task offloading decisions, with the purpose of the whole system reward maximization, i.e., optimizing the global reward. Mathematically, the global reward, denoted by $R^{\text{tot}}$, can be represented as
\begin{equation}
R^{\text{tot}} = \sum_{i=1}^{I} \sum_{t=1}^{T} \gamma \cdot r_i(t)(o_i(t), a_i(t)). 
\end{equation}
However, the global reward $R^{\text{tot}}$ is  obtained only at the end of the episode, resulting in a low training efficiency. To solve this problem, we extract the rewards of all agents in each decision episode to as the instantaneous global reward
\begin{equation}
r^{\text{tot}}(t) = \sum_{i=1}^{I} r_i(t)(o_i(t), a_i(t)). 
\end{equation}
\iffalse
In this way, although ignoring the global reward in a continuous episode duration, we not only provide instantaneous rewards for each agent, thereby improving learning efficiency and enabling faster convergence to the optimal strategy, but also the differences in strategy and performance between the continuous episodes and discrete episodes can be ignored.
\fi
Furthermore, in the context of Dec-POMDP, the scheduling of computing tasks in each decision episode aims to maximize the whole system reward. In addition, the decisions provided by the agents are also effected by each other. Accordingly, above computing task scheduling process can be regarded as an exact potential game (EPG) \cite{GRU1}, which can be defined as a scenario that the change of individual agent's utility by updating its offloading strategy, and be exactly equivalent to the change of all system's potential utilities. 
\iffalse
In other words,1 to some extent, once an agent adjusts its offloading strategy, both its individual utility and the whole system's efficiency are affected.
\fi
With the help of EPG, we conclude that the instantaneous reward maximization of an individual agent can lead to the maximization of the whole system's reward. That is, when the instantaneous reward of $i$-th agent, i.e., $r_i(t)$, is maximized, the total rewards of all agents in each decision episode, i.e. $r^{\text{tot}}(t)$, is also maximized.  Mathematically, the instantaneous global reward can be maximized by updating offloading decisions of all agents as follows
\begin{equation}
\underset{\textbf{a}(t)}{\arg\max} \; r^{\rm tot}(t)(\textbf{o}(t), \textbf{a}(t)) = 
\left(
\begin{aligned}
&\underset{a_1(t)}{\arg\max} \; r_1(t)(o_1(t), a_1(t)) \\
&~~~~~~~~~~~~~~~~\vdots \\
&\underset{a_I(t)}{\arg\max} \; r_I(t)(o_I(t), a_I(t))
\end{aligned}
\right)
\end{equation}
where $\textbf{a}(t) = \{a_1(t), a_2(t), \ldots, a_I(t)\}$ and and $\textbf{o}(t) = \{o_1(t), o_2(t), \ldots, o_I(t)\}$ denotes the set of actions determined by all agents and the set of local spaces observed by all agents at episode $t$, respectively.

Given a local observation of $i$-th agent at episode $t$, i.e., $o_i(t)$, the expected reward when it selects the action $a_i(t)$ can be calculated by the corresponding Q-value function. That is, with the configurations of $o_i(t)$ and $a_i(t)$, the expected reward of $i$-th agent can be represented as
\begin{equation}
\begin{aligned}
Q(o_i(t),a_i(t))= r_i(t) +  \mathbf{E_\pi} \left[\sum_{t'=t+1}^{T} \gamma^{t'-t} r_i(t')\mid o_i(t), a_i(t) \right]
\end{aligned}
\end{equation}
where $\textbf{E}[\cdot]$ is the expectation function, and $\pi$ is a policy rule that guides agents for choosing corresponding actions based on their local observations.

Due to the dynamic nature and uncertainty of the environment, it is difficult to derive the optimal Q-value function for each agent. To solve this problem, we introduce the Bellman equation into the decision-making process of the agents, since it can simplify multi-stage decision problems into a series of single-state problems by using recursive decomposition. In this way, the optimal Q-value function, denoted by $Q^\ast(o,a)$, can be derived, which denotes the maximum expected reward by taking action $a$ within a given local observation $o$ and following the optimal policy, the expression is shown in Eq. \eqref{eq:optimal reward}.
\begin{figure*}[t] % t选项是尽量放置在页面顶部
\begin{equation}\label{eq:optimal reward}
\begin{aligned}
Q^\ast(o_i(t),a_i(t)) = \mathbf{E}[r_i(t) + \gamma\cdot \max_{a_i(t+1)} Q^\ast(o_i(t+1),a_i(t+1)) \mid o_i(t),a_i(t)] 
\end{aligned}
\end{equation}
\hrule % 分割线
\end{figure*}
Furthermore, by implementing a greedy selection mechanism based on the optimal Q-value function of $Q^\ast(o_i(t),a_i(t)$ in Eq.  \eqref{eq:optimal reward}, which prioritizes the action that is anticipated to provide the highest reward in the current state, thereby 
we can derive the corresponding optimal policy for the local observations, which is given by
\begin{equation}\label{eq:optimal policy local}
\pi^\ast = \underset{a_i(t)}{\arg\max} \, Q^\ast(o_i(t),a_i(t)) 
\end{equation}
However, finding the optimal policy by using Eq. \eqref{eq:optimal policy local} ignores the potential long-term benefits of actions that do not be the optimal in the current context, resulting in an inadequate exploration for potential long-term benefits.
To solve this problem, we introduce a $\epsilon$-greedy strategy, under which the agent can either select the best action known currently with the probability of $1-\epsilon$, or choose any feasible action randomly with the probability of $\epsilon$. 

Under the condition of limited strategy space for all agents, the EPG can converge to a pure strategy Nash equilibrium, so the expected maximum cumulative reward can be obtained. However, finding the global optimal action of a task offloading strategy is a challenging problem, due to the dynamic environment in which the agents. 
In the next subsection, we utilize a MADRL method based on CTDE to search for an approximate optimal action for the EPG.

\subsection{CTDE-based DRL Architecture}

\iffalse
\textcolor{red}{With the assumption of limited action space, traditional centralized single DRL methods provided efficient computational resource allocations. However, with the dramatic increase on the number of agents and offloading decisions caused by abundant computing resources, the scale of action space becomes magnificent, resulting in the non-effectiveness of single DRL methods. While for distributed DRL methods, it is difficult to update the parameters among agents because of the vehicle's mobility, affecting the converge of DRL methods. Accordingly, the traditional offloading methods based on centralized DRL and distributed ones are usually offline, the real-timeness and effectiveness of offloading models are affected.
}
\fi

The CTDE, as a cutting-edge framework that integrates a hybrid MARL framework, allows for centralized training of agents using global state information, while enabling agents to act in a decentralized manner based on local observations. However, extracting the policies from the centralized training process that can be decentralized for execution remains a challenge for implementing CTDE in MADRL, since the discrepancy between global information used during training and local observations available during execution leads to issues of policy effectiveness and stability. To overcome this challenge, the VDN algorithm is introduced to solve the problem of aligning decentralized execution with centralized training outcomes, from the perspective of facilitating individual agent decisions that collectively contribute to achieving the same global optimal policy.

Therefore, in this section, based on the framework of CTDE-MADRL, we propose an effective task offloading method empowered by VEH-PLA, called as a heterogeneous computational resources allocation (CTDE-MADRL-HCRA), which integrates the advantages of VEH-PLA coordination, dynamic learning capability of MADRL, efficient collaboration mechanism of CTDE strategies, and decision optimization techniques of VDN to address challenges of online offloading. 

\iffalse
\textcolor{red}{The CTDE-MADRL-HCRA can optimize resource allocation, enhances decision accuracy, reduces processing delay and increases effectiveness of a task offloading in the IoV.}
\fi
\begin{figure}[htbp]
\centerline{\includegraphics[scale=0.38]{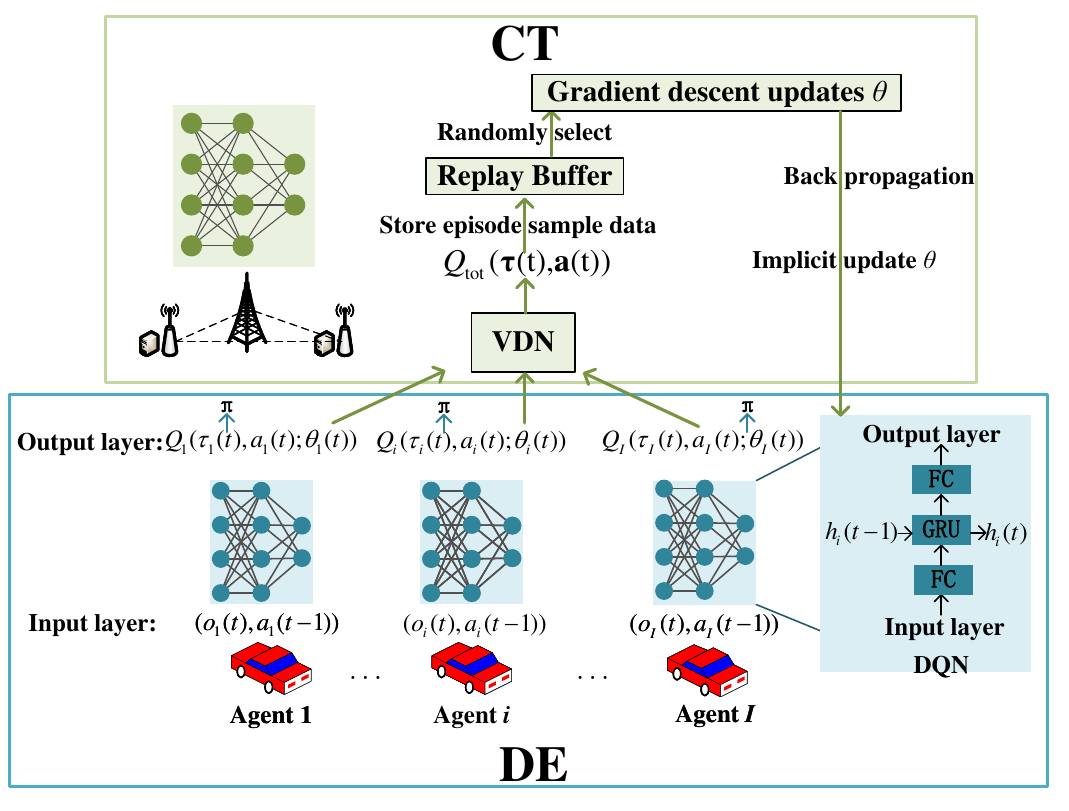}}
\caption{Architecture of the CTDE-based task offloading}
\label{fig:MADRL}
\end{figure}

To narrow the gap between global state information and local state information, the method of storing historical observations as an auxiliary information to construct the Q-value function model has been proven to be effective \cite{GRU1}. Accordingly, we adopt $\bm{\tau}(t) = \{\tau_1(t), \tau_2(t),..., \tau_I(t)\}$ to represent the collection of historical observations of all agents, where $\tau_i(t) = \{(a_i(1),o_i(1),r_i(1)),..., (a_i(t-1),o_i(t-1),r_i(t-1))\}$ is joint action-observation history of $i$-th agent. 

The CTDE-based framework aims to learn a joint Q-value function $Q_{\text{tot}}(\bm{\tau}(t), \bm{a}(t))$ with global state information by using centralized training process. Once $Q_{\text{tot}}(\bm{\tau}(t), \bm{a}(t))$ is learned, which cannot be obtained by using distributed DRL algorithms, distributed agents based on CTDE can make more reasonable independent decisions based on individual local observations. The proposed CTDE-based computational offloading architecture is shown in Fig. \ref{fig:MADRL}, and can address the aforementioned challenges.
\iffalse
\textcolor{red}{i.e., each agent can only make decisions based on its limited local observations, and cannot access the global state information of the entire system, leading to a failure to effectively coordinate actions of the agents to maximize the whole rewards of the system. In addition, the architecture also achieves effective collaboration and decision-making capabilities of the agents in dynamic environments.} 
\fi
It can be noticed that the architecture consists of individual agent modules, a central training mechanism, and a policy execution framework. In detail, individual agent modules are responsible for processing local observations and selecting corresponding actions based on their current policies. The central training mechanism can solve the challenge of learning a joint Q-value function that incorporates global state information, enabling the optimization of collective outcomes. The policy execution framework makes it possible for agents to act independently in a decentralized manner during execution, relying on their local observations while still aligning their actions with the learned global objectives.

%\textcolor{blue}{Based on the VDN technology, we can decompose the joint Q-value function $Q_{\text{tot}}(\bm{\tau}(t), \bm{a}(t))$ into the Q-value functions of individual agent, denoted by $Q_i(\tau_i(t), a_i(t); \theta_i(t)), i \in I$, where $\theta_i(t)$ is the parameter of $i$-th agent in corresponding neural Q network. That is, \begin{equation}\label{eq:decompose q-value}Q_{\text{tot}}(\bm{\tau}(t), \bm{a}(t)) \approx \sum_{i \in I} Q_i(\tau_i(t), a_i(t); \theta_i(t)).\end{equation}}

To achieve the decentralized executions executed from centralized training outcomes, based on the VDN technology, we can decompose the joint Q-value function $Q_{\text{tot}}(\bm{\tau}(t), \bm{a}(t))$ at episode $t$ into the Q-value functions of individual agent, denoted by $Q_i(\tau_i(t), a_i(t)), i \in I$. That is, 
\begin{equation}\label{eq:Qtot}
Q_{\text{tot}}(\bm{\tau}(t), \bm{a}(t)) \approx \sum_{i \in I} Q_i(\tau_i(t), a_i(t)).
\end{equation}
In practical application,  constructing or maintaining directly a Q-value table for each state-action pair is impractical, as the state space gets very large with the increasing of agents, making the Q-value function difficult to be stored or computed directly. Therefore, we use a deep Q-value network (DQN) to approximate the above function, that is, we adopt $Q_i(\tau_i(t), a(t)) \approx Q_i(\tau_i(t), a_i(t); \theta_i(t))$, where $\theta_i(t)$ is the parameter of $i$-th agent's Q-value network.
Therefore, Eq. \eqref{eq:Qtot} can be rewritten as
\begin{equation}
\label{eq:decompose q-value}Q_{\text{tot}}(\bm{\tau}(t), \bm{a}(t)) \approx \sum_{i \in I} Q_i(\tau_i(t), a_i(t); \theta_i(t)).
\end{equation}

In this way, the decomposition from joint Q-value function to that of individual agent can be simplified for the learning task.
Furthermore, with the help of VDN technique, the joint Q-value function $Q_{\text{tot}}(\bm{\tau}(t), \bm{a}(t))$ obtained from centralized learning can be embedded in the distributed execution process. In particular, to maintain the consistency in decision-making process, it is necessary to ensure that the results obtained by globally optimizing the joint Q-value function $Q_{\text{tot}}(\bm{\tau}(t), \bm{a}(t))$ (i.e., the global argmax operation) are consistent with the results set obtained by independently optimizing the Q-value function $Q_i(\tau_i(t), a_i(t))$ for each $i$-th agent, i.e.,
\begin{equation}
\arg\max_{\bm{a}(t)} Q_{\text{tot}}(\bm{\tau}(t), \bm{a}(t)) = 
\left(
\begin{array}{c}
\underset{a_1(t)}{\arg\max} Q_1(\tau_1(t), a_1(t)) \\
\vdots \\
\underset{a_I(t)}{\arg\max} Q_I(\tau_I(t), a_I(t))
\end{array}
\right)
\end{equation}
\iffalse
Therefore, in distributed execution of CTDE, based on local observations and independent decisions, the behavior of all agents can still effectively guarantee the system convergence towards the global optimum.
\fi
To facilitate the agents' learning from their interaction histories and enhance decision-making accuracy, the historical observations $\tau_i(t)$ can be stored in a Gated Recurrent Unit (GRU) and serve as a valuable supplementary information for enhancing the agent's Q-value function model. By minimizing the difference between the joint target Q-values and their current joint Q-values, we can obtain the optimal joint Q-value function. Let $Q_i^-$ be the target Q-value of $i$-th agent. Based on the result of Eq. \eqref{eq:decompose q-value}, we get $Q_{\text{tot}}^-(\bm{\tau}(t), \bm{a}(t)) \approx \sum_{i \in I} Q_i^-(\tau_i(t), a_i(t); \theta_i(t))$. Accordingly, the loss function of \(Q_{\text{tot}}\) is defined as
\begin{equation}
L(\theta) = \frac{1}{X} \sum_{x=1}^{X} \sigma_x^2 = \frac{1}{X} \sum_{x=1}^{X} (y_{\text{tot}}^x - Q_{\text{tot}}^x(\tau^x, a^x))^2
\label{L}
\end{equation}
where $\sigma$ is the difference between a difference target $y_{\text{tot}}$ and a joint Q-value $Q_{\text{tot}}$. The temporal difference target is an estimate by an agent of future rewards obtainable from current state and action, denoted as \(y_{\text{tot}}^x(t) = r^x(t) + \gamma \max_{a(t+1)} Q_{\text{tot}}^-(\bm{\tau}(t+1), \bm{a}(t+1))\), $X$ is number of samples.

The loss function of the joint Q-value function in Eq. \eqref{L} cannot properly deal with the gradient estimation, since the interaction among multiple agents leads to a non-stationary target for optimization. With the help of the back propagation method, we can obtain an implicit update of individual agent networks based on global rewards rather than single reward of each agent. Furthermore, by updating individual Q-value functions of the agents and sharing the weights of the agent networks, the training complexity can be further simplified. To sum up, the pseudo-code of CTDE-MADRL-HCRA is shown in Algorithm \ref{alg:task_offloading}, where $T_{\max}$ denotes the maximum number of epochs in the whole training, $t_{\max}$ is the maximum number of episodes in each epoch. Within each episode, all agents engage in a complete interaction with the environment. Following each epoch, sample collection and gradient updates are performed.

\begin{algorithm}
\caption{CTDE-MADRL-HCRA Algorithm}
\label{alg:task_offloading}
\begin{algorithmic}[1]
\Input The local observation spaces of all agents $\{O_i(t)\}_{i \in I}$, \Statex \hspace{12pt} and action spaces available for all agents $\{A_i(t)\}_{i \in I}$
\Output The task offloading decisions of all agents $\Psi(t)$
\For{$T = 1$ to $T_{\max}$}
  \State The $i$-th agent obtains the local observation of state \Statex \hspace{14pt} information, i.e., $o_i(t)$
    \For{$t = 1$ to $t_{\max}$}
        \For{$i$-th agent ($1\leq i \leq I$)}
            \If{the corresponding task is public and has been \Statex \hspace{50pt} offloaded by other agents}
                \State Skipping offloading
            \Else
                \State Select a random action with the probability \Statex \hspace{56pt} of $\epsilon$ for exploration, select $a_i(t) =$\Statex \hspace{56pt} $\underset{a_i(t)}{\mathrm{argmax}}\ Q(o_i(t),a_i(t);\theta_i)$ for exploitation \Statex \hspace{56pt}with the probability of $1-\epsilon$
            \EndIf
        \EndFor
        \State Execute joint action $\bm{a}(t)$, obtain joint reward \Statex \hspace{26pt} $\bm{r}(t) = \{r_1(t), r_2(t), ..., r_I(t)\}$, new local \Statex \hspace{26pt} observations $\bm{o}(t+1)$, and global state $s(t+1)$
        \State Store data samples of $\bm{o}(t), \bm{a}(t), \bm{r}(t), \bm{o}(t+1)$ in \Statex \hspace{26pt} the buffer
    \EndFor
    \State Randomly draw a batch of samples from the buffer
    \State Update $\theta$ using gradient descent based on Eq. \eqref{L}
    \State Update the target Q-network parameters $\theta^{-}$ of each \Statex \hspace{12pt} agent using $\theta$
\EndFor
\end{algorithmic}
\end{algorithm}

\subsection{Convergence and Complexity Analysis}\label{sub:Initialization, Convergence, and Complexity}
\iffalse
\subsubsection{Initialization}
In order to facilitate an efficient start to the CTDE-MADRL-HCRA Algorithm, we investigate to properly initialize the predictive Q-value networks for all agents. The predictive Q-value network of each agent is established according to the parameters $\theta$, while the corresponding target Q-value network parameters are set to $\theta^-$ and are initialized to $\theta$. In this way, parameter alignment operation can ensure that all agents start from a common baseline, allowing for an accurate evaluation of task offloading decisions within a shared framework. The synchronization of the predictive and target Q-networks forms the foundation for the collective learning and decision-making process of the agents.
\fi

\subsubsection{Pre-training for Algorithm Convergence}
Considering the real-world IoV scenarios, there could be hundreds of vehicles/agents per kilometer of road, and the number of vehicles may fluctuate over time. Therefore, achieving a stable policy convergence requires a considerable amount of episodes. To address this issue, the collaboration of multiple powerful edge servers for pre-training is considered. Based on empirical statistical methods, setting up a large-scale number of agents beforehand can achieve an approximate convergence of the policy. In practical applications, if the number of agents does not reach the scale setting beforehand, the environmental and action information of the redundant agents is set to null. When new agents join the scene, their information can be directly filled into the preset agents with no actions, effectively adapting to the complexity and dynamism of the VEC network and significantly improving the efficiency of the decision-making process and the overall performance of the system.
%In the actual deployment of IoV, especially considering the scenario where hundreds of ICV may crowd every kilometer of road and the number of ICV fluctuates over time, the training time required for the Policy Agent (PA) to achieve stable policy convergence presents a significant challenge. To address this issue, this study introduces a solution: pre-training conducted on powerful remote cloud servers or collaboratively across several edge servers. This process, based on empirical statistical methods and aimed at a large-scale preset number of ICVs, is designed to quickly achieve approximate convergence of policies.

%Furthermore, to accommodate the variability in the number of ICV in online scenarios, our strategy applies well-trained models to these settings. It integrates new state and action variables into the environment as they arise, thus flexibly adapting to the dynamic changes in the number of ICV. The core of this strategy is the adoption of a CTDE algorithm, conducted on a master RSU server with ample resources. This method not only effectively adapts to the complexity and dynamics of the VEC network but also significantly improves the efficiency of the decision-making process and overall system performance.

%\subsection{Time Complexity Analysis}
%\textcolor{red}{xxxx.}
\subsubsection{Complexity Analysis}
%在我们的研究中，由于集中训练阶段发生在整个系统的准备阶段，因此主要关注的是分布式执行的计算复杂度。通过采用在每个车联网上部署独立神经网络的策略，实现了异构车联网环境下的并行执行能力。这种方法允许分析，最初针对多智能体系统的整体计算复杂性，现在集中在单个ICV上执行单个神经网络的复杂性。
The centralized training of CTDE can be achieved before the beginning of CTDE-MADRL-HCRA, then corresponding complexity depends on the distributed execution of CTDE. 
\iffalse
Furthermore, by employing a strategy of deploying independent neural networks on each agent, parallel execution capabilities in a heterogeneous IoV environment are achieved. In this way, we can focus on the computational complexity of executing a single neural network of an individual agent instead of the overall complexity of multi-agent system.
\fi
Let $P_1$, $P_2$, and $P_3$ denote the number of neurons in the input layer, hidden layer and output layer, respectively, and $V$ denote the sequence length of historical observations, which can affect the computation of the GRU layer belonging to the hidden layer. In line 8 of CTDE-MADRL-HCRA, the input layer, without involving any data processing, is responsible for transferring the input data to the hidden layer, corresponding computational complexity can be ignored. Furthermore, the hidden layer includes fully connected layer and GRU layer for solving Q-value function. Computational complexity of the fully connected layer is $O(P_1P_2)$, since each one of output neurons needs to be connected to and compute with every neuron in the previous layer. Based on the assumption that each unit within the GRU interacts with other units at each timestep, the computational complexity of the GRU layer is $O(VP_2^2)$. The computational complexity of the output layer is $O(P_2 P_3)$, since it not only is a fully connected layer, but also is tasked with transforming the output of the GRU layer into an action policy. To sum up, considering the computational complexity caused by input layer, hidden layer and output layer, overall computational complexity of CTDE-MADRL-HCRA is $O(P_2(P_1 + VP_2 + P_3))$.

\section{Evaluations}\label{sec:Simulation Results}
\subsection{Simulation Sep}
\begin{table}[H]
\caption{The Settings of Simulation Parameters}
\label{tab:simulation_parameters}
\centering
\setlength{\tabcolsep}{3pt} % 减少列间距
\renewcommand{\arraystretch}{1.3} % 减少行高
\begin{tabular}{>{\centering\arraybackslash}m{0.7cm} >{\centering\arraybackslash}m{5.6cm} >{\centering\arraybackslash}m{1.8cm}}
\Xhline{2\arrayrulewidth}
Notation & Definition & Value \\
\hline
$K$ & Number of VEH-PLA & 2 \\
$m$ & Number of VEH-PLA members & 5 \\
\(N\) & Number of vehicles connected to a RSU & 6 \\
$L$ & Coverage of each RSU & 400 m \\
\(P_{i}\) & Transmission power of each vehicle & 20 dBm\textsuperscript{\cite{10147826}} \\
\(\mu\) & Mean of the truncated Gaussian probability distribution & 60\textsuperscript{\cite{10147826}} \\
\(\sigma\) & Standard deviation of the truncated Gaussian probability distribution & 5\textsuperscript{\cite{10147826}} \\
\(B\) & Bandwidth of sub-6G band & 30 MHz\textsuperscript{\cite{10147826}} \\
\(\theta\) & Shadowing standard deviation & 3.5 dB \textsuperscript{\cite{10147826}} \\
\(p_{0}\) & Channel power gain per unit distance & -65 dB\textsuperscript{\cite{10147826}} \\
\(F^{V}_{\max}\) & Computing resource of a vehicle & 0.5 GHz\textsuperscript{\cite{10147826}} \\
\(F^{R}_{\max}\) & Computing resource of a RSU & 6 GHz\textsuperscript{\cite{10147826}} \\
\(F^{M}_{\max}\) & Computing resource of a MBS & 10 GHz\textsuperscript{\cite{10147826}} \\
\(\lambda\) & Task bit size & 1.5 Mbits \\
\(\rho\) & Task computation density & 100 cycles/bit \\
\(\gamma\) & Discount factor & 0.9 \\
\(\alpha\) & Learning rate & 0.0003 \\
\(\epsilon\) & Exploration factor & [0.05, 1] \\
\Xhline{2\arrayrulewidth}
\end{tabular}
\end{table}
In this section, we validate the effectiveness and correctness of the proposed method through Python 3.11 and TensorFlow 2.14.0 simulator. In detail, a bi-directional lane with the length of 1200 meters is constructed, and the vehicles' position are reset to the starting point when they reach the end of the road for avoiding offloading disruptions caused by the vehicles leaving the environment.
Three RSUs are fixed at the position of 200m, 600m, and 1000m, and 15 moving vehicles are randomly distributed. To reduce the size of the action space and achieve load balancing, if the number of vehicles served by the same RSU is above $N$, the RSU will no longer provide the sources to other vehicles with the demand of task offloading, which can be transferred to the adjacent RSUs. In addition, the related simulation parameters are given in Table~\ref{tab:simulation_parameters}. 
In addition, to model the task categorization (i.e., public tasks and personalized tasks) for the VEH-PLA members and non-platooning members, the ratio of public tasks for the latter is 20\%, while that of VEH-PLA members is set to 30\%, since the vehicles in VEH-PLA usually have similar task requirements.

\subsection{Parameter Settings of DRL Model}
In the training processing of the DRL model, the discount factor \(\gamma\) is set to 0.9 for reflecting the relative importance of future rewards to immediate ones. The learning rate \(\alpha\) is determined to be 0.0003 to regulate the pace of model weight updates, ensuring stability and convergence speed in the learning progression. The exploration factor \(\epsilon\) is initially set to 1 and decays exponentially with an increase in exploration attempts, diminishing ultimately to 0.05. These settings allow the model to extensively explore potential action strategies initially and gradually focus on enhancing the best strategies discovered. A buffer with the size of 2000 provides the learning model with an ample experiential reservoir. An update interval for the target, occurring once every 100 episodes, contributes to the stability of target values during learning. Furthermore, the mini-batch size is set to 64, optimizing the efficiency of the training process and data utilization.

\subsection{Baseline Schemes}

To validate the effectiveness and correctness of task offloading decisions based on VEH-PLA proposed in the paper, i.e., \textbf{CTDE-MADRL-HCRA}, which is based on VDN, VEH-PLA and task categorization, we adopt offloading baseline schemes based on VDN, IQL and QMIX, respectively. In particular, with the consideration of the VEH-PLA and task type or not, baseline schemes can be  classified as follows. 

\begin{itemize}
   \item \cite{10147826}: Offloading schemes based on the VDN, and these based on VEH-PLA and task categorization, denoted by \textbf{VDN-NoVP-NoTC},  \textbf{VDN-VEH-PLA} and \textbf{VDN-TC}; 
  \item \cite{rashid2020monotonic}: Offloading schemes based on the QMIX, and along with both VEH-PLA and task categorization, denoted by \textbf{QMIX-NoVP-NoTC}, \textbf{QMIX-VEH-PLA-TC}; 
  \item \cite{tampuu2017multiagent}: Offloading schemes based on the IQL, and along with both VEH-PLA and task categorization, denoted by \textbf{IQL-NoVP-NoTC} and \textbf{IQL-VEH-PLA-TC}; 
  \item \cite{local}: \textbf{LOCAL} offloading;
  \item \cite{LRSU}: \textbf{LOCAL and RSU} offloading.
\end{itemize}

In the following, we investigate the impact of the numbers of episodes and vehicles, the sizes of VEH-PLA and task data on the convergence rate, average reward, and processing delay.

\begin{table}[ht]
\centering
\caption{The comparison of average cumulative rewards}
\label{tab:accumulated_reward}
\begin{tabularx}{\linewidth}{@{}Xlll@{}}
\toprule
Scheme Name & Acc. Reward & Diff. Previous & Diff. First \\
\midrule
CTDE-MADRL-HCRA & 95.562 & --- & ---\\
VDN-NoVP-NoTC \textsuperscript{\cite{10147826}}  & 84.673 & 10.23\% ↑& 12.86\% ↑ \\
IQL-VEH-PLA-TC \textsuperscript{\cite{tampuu2017multiagent}}       & 91.298 & --- & 04.96\% ↑ \\
IQL-NoVP-NoTC \textsuperscript{\cite{tampuu2017multiagent}}  & 83.646 & 09.15\% ↑ & 14.25\% ↑ \\
\bottomrule
\end{tabularx}
\end{table}

\begin{figure}[htbp]
  \centering % 让图片居中显示
  \includegraphics[width=0.32\textwidth]{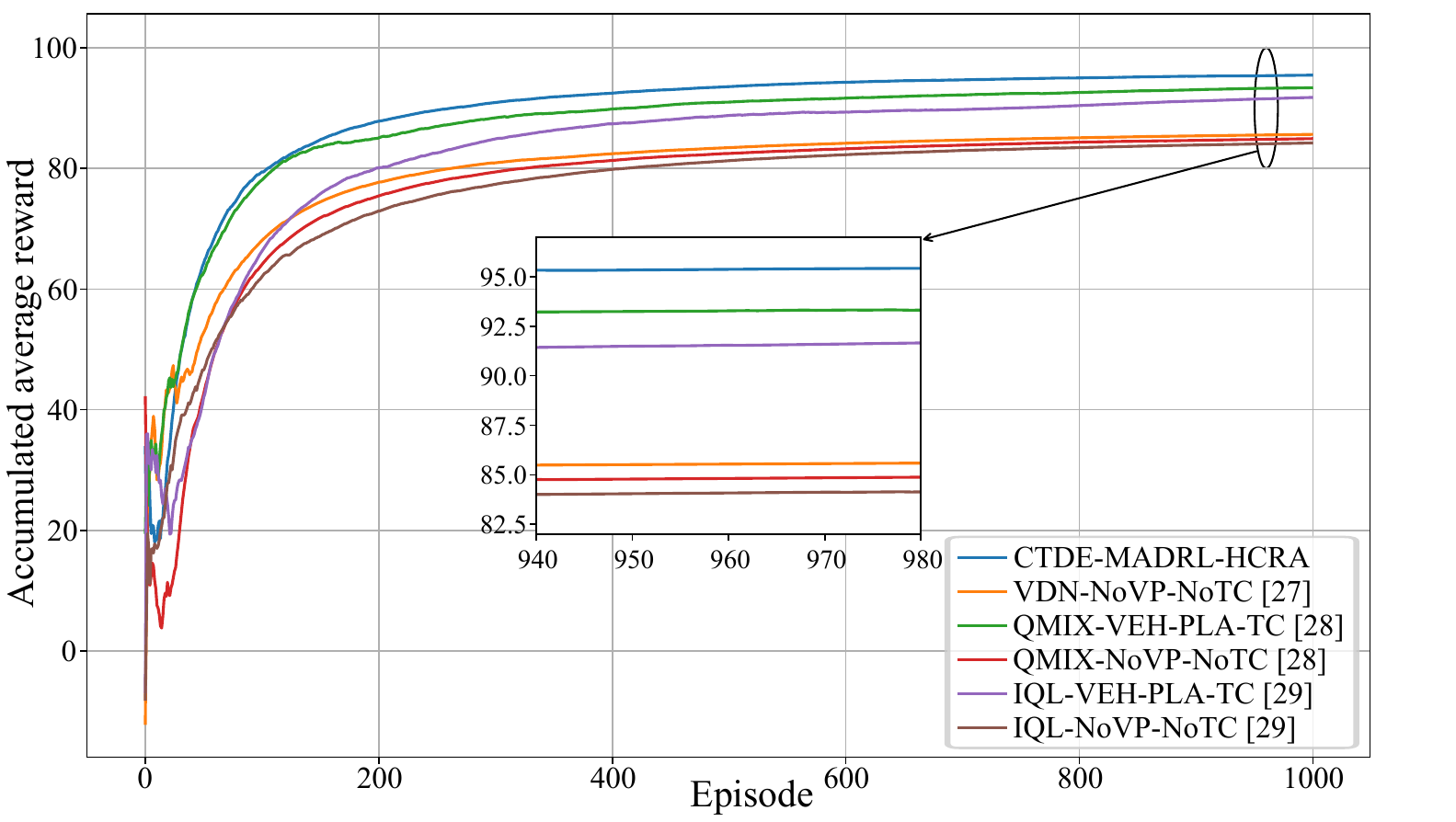} % 图片文件的路径
  \caption{Episode vs. average accumulated reward  } % 图片的标题
  \label{fig:Accumulated Average Reward} % 用于引用图片的标签
\end{figure}

With the settings of 15 vehicles, consisting of two VEH-PLA with 5 vehicles respectively, and 5 non-platooning vehicles, other simulated parameters are set based on Table \ref{tab:simulation_parameters}, Fig. \ref{fig:Accumulated Average Reward} demonstrates the relationship between the accumulated average reward nd the number of training episodes. On the one hand, it can be noticed that, with the increment of the number of training episodes, \textbf{CTDE-MADRL-HCPA}, along with other baseline schemes, shows a fast convergence speed and good stability, which validates the effectiveness analyzed in Subsection \ref{sub:Initialization, Convergence, and Complexity}. On the other hand, the average rewards of \textbf{CTDE-MADRL-HCRA}, as well as \textbf{QMIX-VEH-PLA-TC} and \textbf{IQL-VEH-PLA-TC} employing the offloading model proposed in this paper, are significantly higher than these of baseline schemes. The corresponding reason is that the VEH-PLA and task categorization reduce the number of task offloading requests, and further reduce computational resource consumption of vehicles belonging to VEH-PLA, which can be used to satisfy the demand of task offloading requiring by other vehicles. In particular, as shown in Table ~\ref{tab:accumulated_reward}, compared with \textbf{VDN-NoVP-NoTC} \cite{10147826}, \textbf{QMIX-VEH-PLA-TC} \cite{rashid2020monotonic}, \textbf{QMIX-NoVP-NoTC} \cite{rashid2020monotonic}, \textbf{IQL-VEH-PLA-TC} \cite{tampuu2017multiagent}, and \textbf{IQL-NoVP-NoTC} \cite{tampuu2017multiagent},  average reward achieved by \textbf{CTDE-MADRL-HCPA} is increased by 11.45\%, 2.38\%, 12.86\%, 4.96\% and 14.25\%, respectively, which 
suggests that the \textbf{CTDE-MADRL-HCRA} possesses a superior
offloading decision optimization capabilities by using task categorization and VEH-PLA, making it particularly suitable for
IoV systems that demand high stability and efficiency.

\begin{figure}[htbp]
  \centering % 让图片居中显示
  \begin{subfigure}[b]{0.32\textwidth}
    \includegraphics[width=\textwidth]{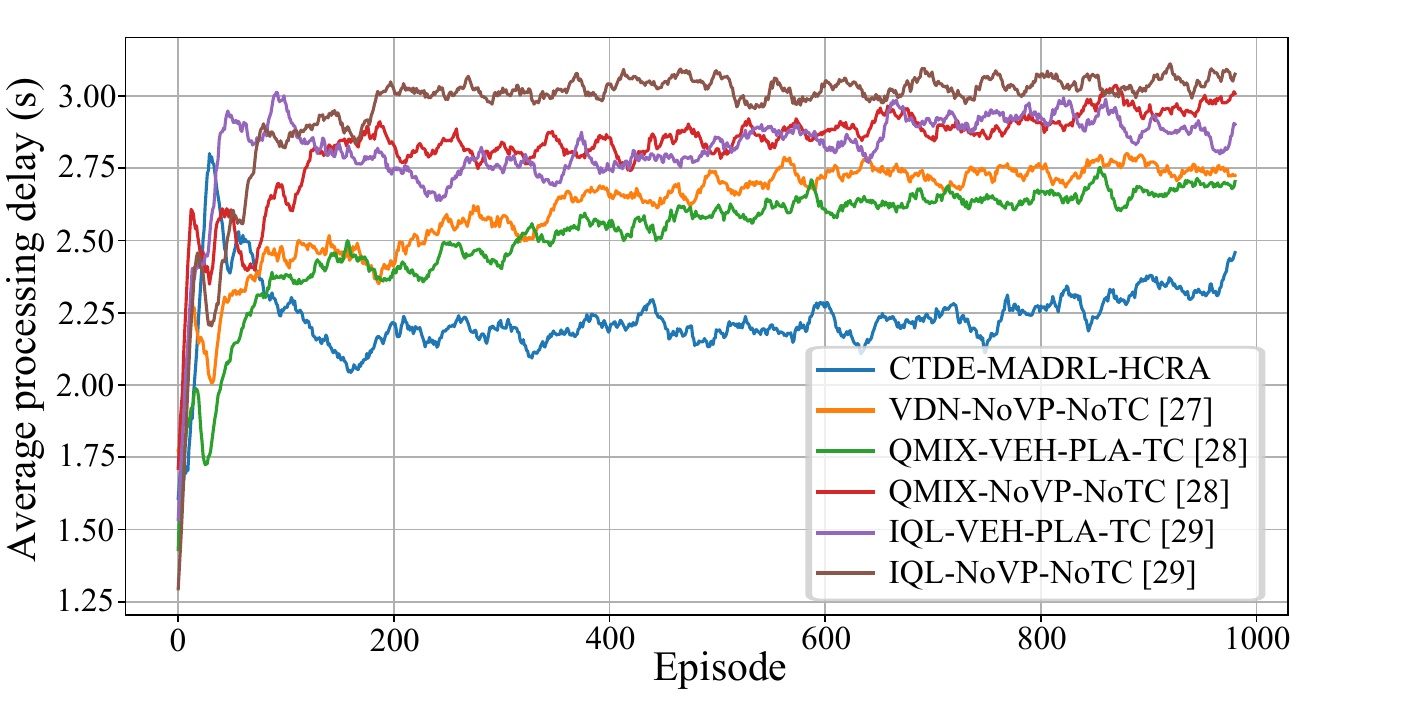} % 图片文件的路径
    \caption{Episode vs. average processing delay (1)} % 子图(a)的标题
    \label{fig:latency-a} % 用于引用子图(a)的标签
  \end{subfigure}
  \hfill % 如果需要的话，可以在子图之间添加一些水平空间
  \begin{subfigure}[b]{0.32\textwidth}
    \includegraphics[width=\textwidth]{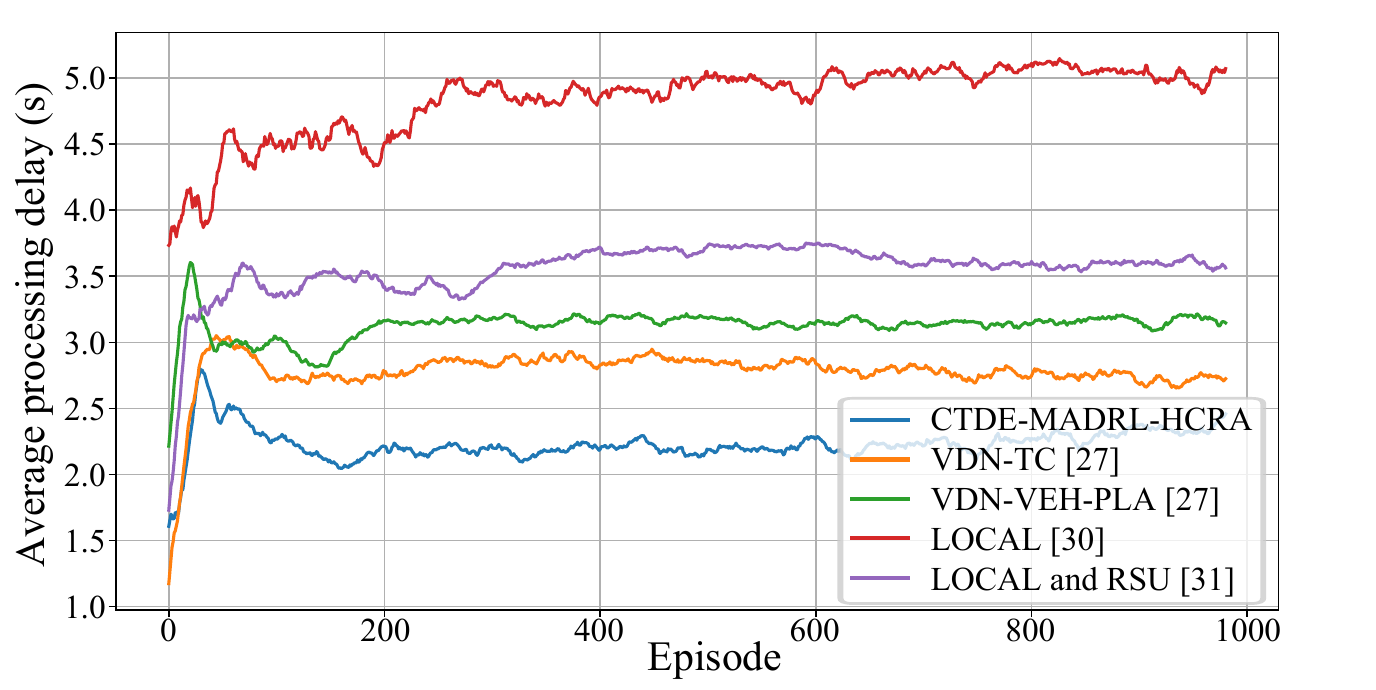} % 第二个图片文件的路径
    \caption{Episode vs. average processing delay (2)} % 子图(b)的标题
    \label{fig:latency-b} % 用于引用子图(b)的标签
  \end{subfigure}
  \caption{Episode vs. average processing delay  } % 总的图形标题
  \label{fig:latency} % 用于引用整个图的标签
\end{figure}

Next, with the same parameter settings of Fig. \ref{fig:Accumulated Average Reward}, average processing delay of task offloading is considered, since the processing delay in Eq. \eqref{objective} cannot fully reflect the exploratory nature of individual vehicle decisions within the processing of DRL, while the average processing delay can smooth out this episodic variability, offering a more consistent measure for assessing the overall system efficiency. Consequently, in Fig. \ref{fig:latency}, we observe the influence of the number of episodes on the average processing delay. It can be observed that, apart from \textbf{LOCAL} offloading scheme \cite{local}, the average processing delay is relatively low in the initial phase of other schemes. This is because that, in the initial phase, the computational resources of RSUs and vehicles in VEH-PLA are not occupied by vehicles requesting for task offloading, and thus have abundant computational resources. Accordingly, the tasks can be offloaded directly successfully. While in \textbf{LOCAL} offloading scheme, the computational resource of individual vehicle is less than these of other offloading schemes, resulting in a larger average processing delay.
However, as the number of episodes increases, more task offloading requests exist and available computational resources become less, resulting in a gradual rise in average processing delay. Among all schemes, \textbf{CTDE-MADRL-HCRA} has the lowest average processing delay followed by the \textbf{QMIX-VEH-PLA-TC} scheme \cite{rashid2020monotonic}, which further demonstrates the effectiveness of proposed offloading model based on EH-PLA. On the average, compared with other offloading schemes \textbf{VDN-NoVP-NoTC} \cite{10147826}, \textbf{QMIX-VEH-PLA-TC} \cite{rashid2020monotonic}, \textbf{IQL-VEH-PLA-TC}  \cite{tampuu2017multiagent}, the processing delay achieved by \textbf{CTDE-MADRL-HCRA} is decreased by 14.52\%, 6.46\% and 13.52\% at least, respectively. In addition, compared to the offloading schemes based on IQL (i.e.,  \textbf{IQL-VEH-PLA-TC}, \textbf{IQL-NoVP-NoTC} \cite{tampuu2017multiagent}), which independent updating policies of multiple agents lead to difficulties in coordinating the optimization of average processing delay, both \textbf{CTDE-MADRL-HCRA} and \textbf{QMIX-VEH-PLA-TC} are more suitable for cooperative environments and therefore exhibit a more better performance.

\begin{figure}[htbp]
  \centering % 让图片居中显示
  \includegraphics[width=0.32\textwidth]{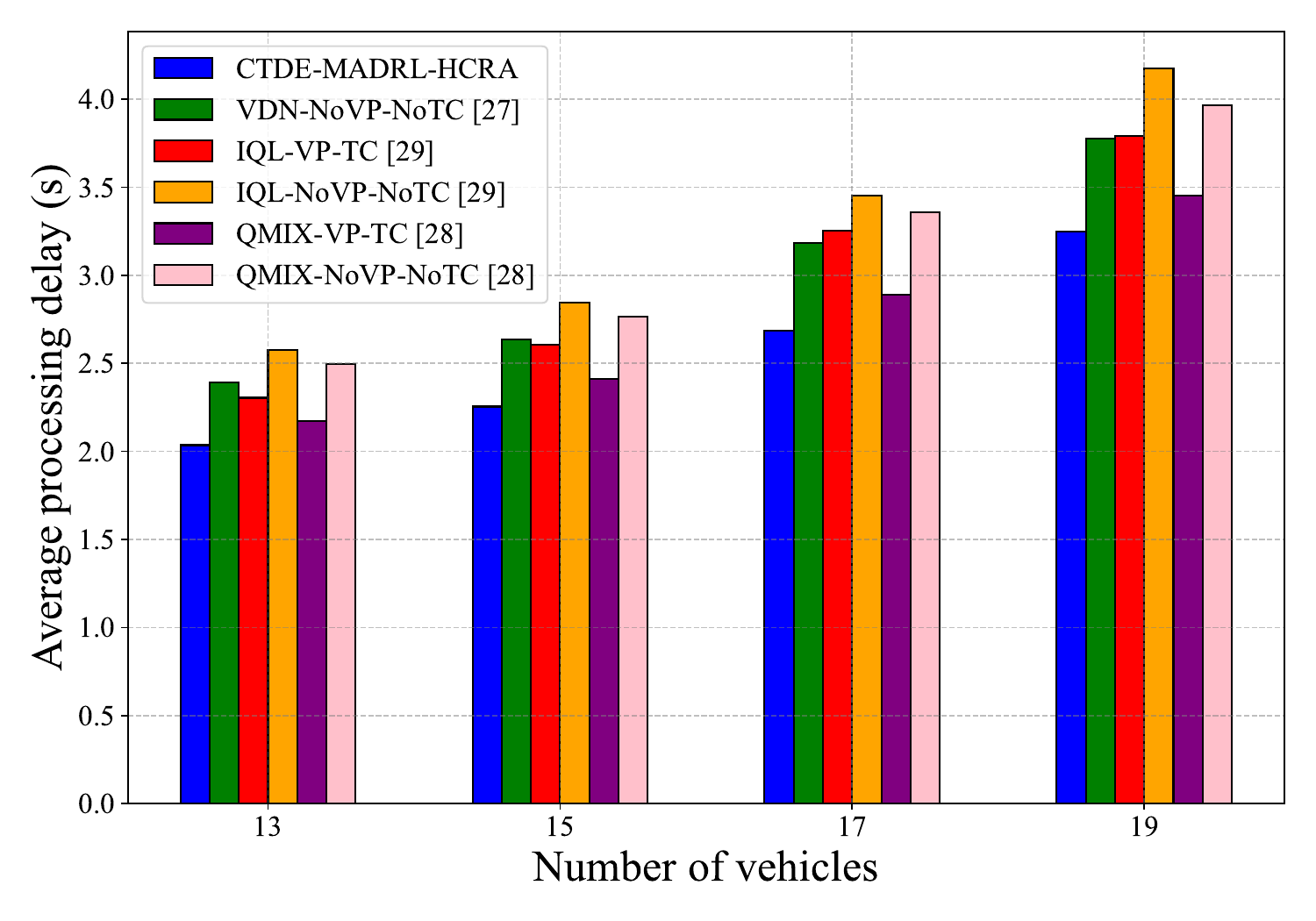} % 图片文件的路径
  \caption{Number of vehicles vs. average processing delay} % 图片的标题
  \label{fig:车数} % 用于引用图片的标签
\end{figure}

\begin{table}[ht]
\centering
\caption{Average processing delay vs. number of vehicles}
\label{tab:system_latency_improvement}
\begin{tabularx}{\linewidth}{@{}Xlllll@{}}
\toprule
Schemes & 13 to 15 & 15 to 17 & 17 to 19 & 13 to 19 \\
\midrule
CTDE-MADRL-HCRA & 10.76\% & 19.07\% & 20.95\% & 59.52\% \\
VDN-NoVP-NoTC \textsuperscript{\cite{10147826}} & 10.37\% & 20.73\% & 18.59\% & 58.02\% \\
QMIX-VEH-PLA-TC \textsuperscript{\cite{rashid2020monotonic}} & 11.01\% & 19.89\% & 19.52\% & 59.06\% \\
QMIX-NoVP-NoTC \textsuperscript{\cite{rashid2020monotonic}}  & 10.81\% & 21.33\% & 18.13\% & 58.82\% \\
IQL-VEH-PLA-TC \textsuperscript{\cite{tampuu2017multiagent}} & 13.13\% & 24.85\% & 16.54\% & 64.61\% \\
IQL-NoVP-NoTC \textsuperscript{\cite{tampuu2017multiagent}} & 10.59\% & 21.29\% & 20.93\% & 62.20\% \\
\bottomrule
\end{tabularx}
\end{table}

Furthermore, for the better observation, the results are averaged with the number of episodes varying from 500 to 1000. Considering different number of non-platooning vehicles and other simulated parameter settings keep the same as before, Fig. \ref{fig:车数} illustrates the performance comparison of all offloading schemes in terms of the average processing delay. It can be noticed that from Fig. \ref{fig:车数} the average processing delay increases with the increment of the number of vehicles. On the one hand, \textbf{CTDE-MADRL-HCRA} shows the best performance with the consideration of VEH-PLA and tesk categorization. On the other hand, as shown in Table ~\ref{tab:system_latency_improvement}, compared with \textbf{CTDE-MADRL-HCRA} and \textbf{QMIX-VEH-PLA-TC}, the growth rate of \textbf{IQL-VEH-PLA-TC} is about 5\%. Compared with offloading schemes based on IQL, such as \textbf{IQL-VEH-PLA-TC} \cite{tampuu2017multiagent}, average processing delays achieved by \textbf{CTDE-MADRL-HCRA} and these based on \textbf{QMIX-VEH-PLA-TC} \cite{rashid2020monotonic} increase more stably, which can be attributed to the fact that, for offloading schemes based on the IQL, the vehicles only learn independently without cooperation. When the number of vehicles becomes large, offloading schemes based on VDN and QMIX perform more effectively.

\begin{figure}[htbp]
  \centering % 让图片居中显示
  \includegraphics[width=0.32\textwidth]{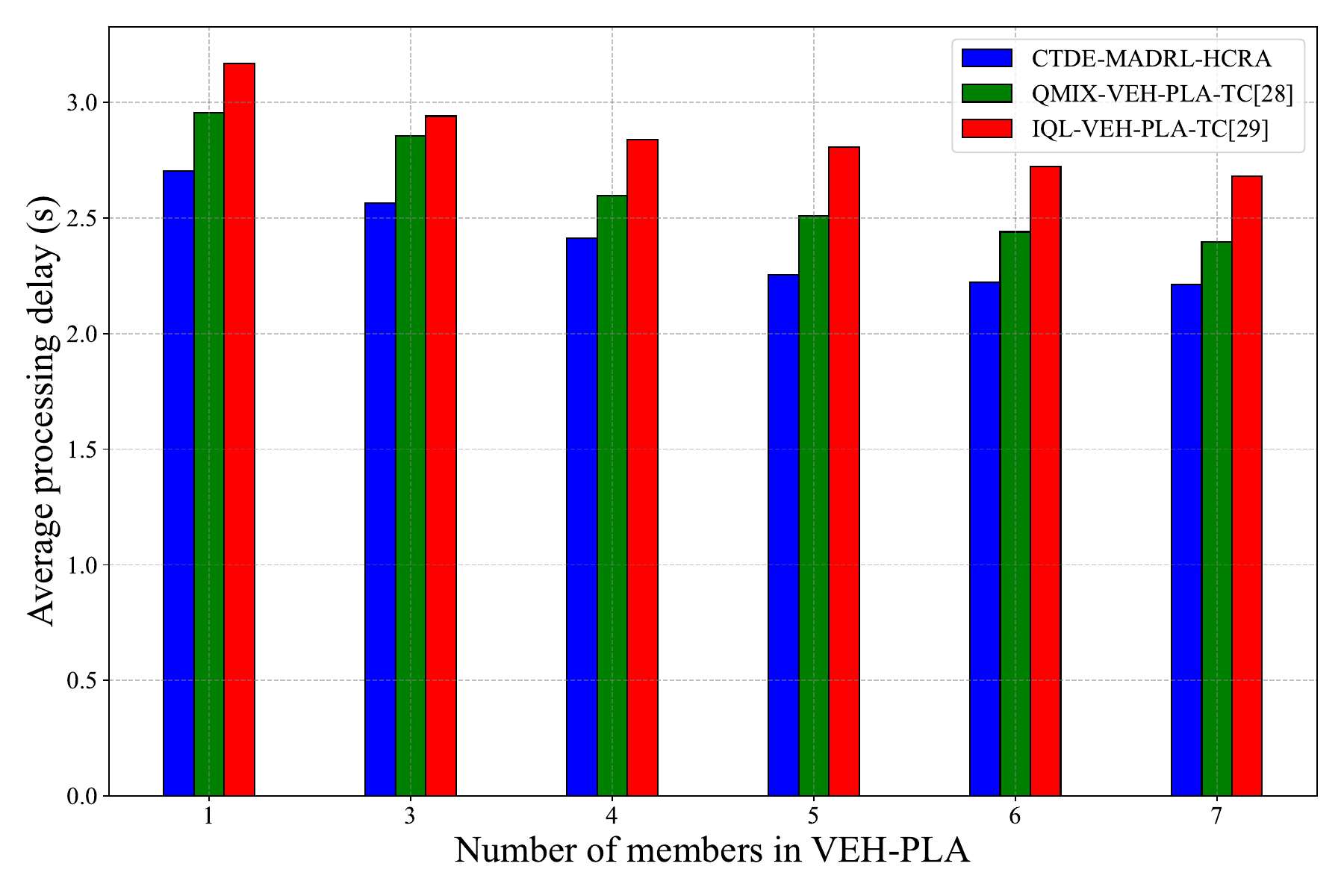} % 图片文件的路径
  \caption{Members in VEH-PLA vs. average processing delay} % 图片的标题
  \label{fig:Number of Platooning Members} % 用于引用图片的标签
\end{figure}

%For the better observation, the results are averaged with the number of episodes varying from 500 to 1000. 
Considering the case of two VEH-PLAs with the number of moving vehicles varying in the range of [3, 4, 5, 6, 7], and the case of no-VEH-PLA (i.e., the number of vehicles in VEH-PLA is 1), Fig. \ref{fig:Number of Platooning Members} illustrates the relationship between the size of VEH-PLA and average processing delay, where the number of vehicles represented at x-axis is that of a single VEH-PLA. It can be noticed that with the increment of VEH-PLA members, on the one hand, the average processing delay shows a downward trend. This is because the fact that over the size of VEH-PLA increasing, the vehicles with the demand of public task offloading can directly utilize the results of the VEH-PLA, which can avoid overlapping offloading for the same public task, decreasing the average processing delay.
On the other hand, total decrement of average processing delay between any two number of VEH-PLA members becomes less significant. In the \textbf{CTDE-MADRL-HCRA} scheme, the average processing delays with the size of VEH-PLA being 5, 6, 7 are 2.254s, 2.223s, and 2.211s, respectively. Compared with average processing delays achieved by 5 and 6 VEH-PLA members, that of VEH-PLA with 7 members is decreased 1.44\% and 1.91\%. One possible is that the marginal benefit of adding additional vehicles to the VEH-PLA diminishes as the number of vehicles increases, indicating a plateau in efficiency gains. Therefore, in the case of Fig. \ref{fig:Number of Platooning Members}, maintaining a VEH-PLA with the size of 5 or 6 members is a more appropriate choice, in terms of average processing delay, VEH-PLA stability and offloading feasibility.

\begin{figure}[htbp]
  \centering % 让图片居中显示
  \includegraphics[width=0.32\textwidth]{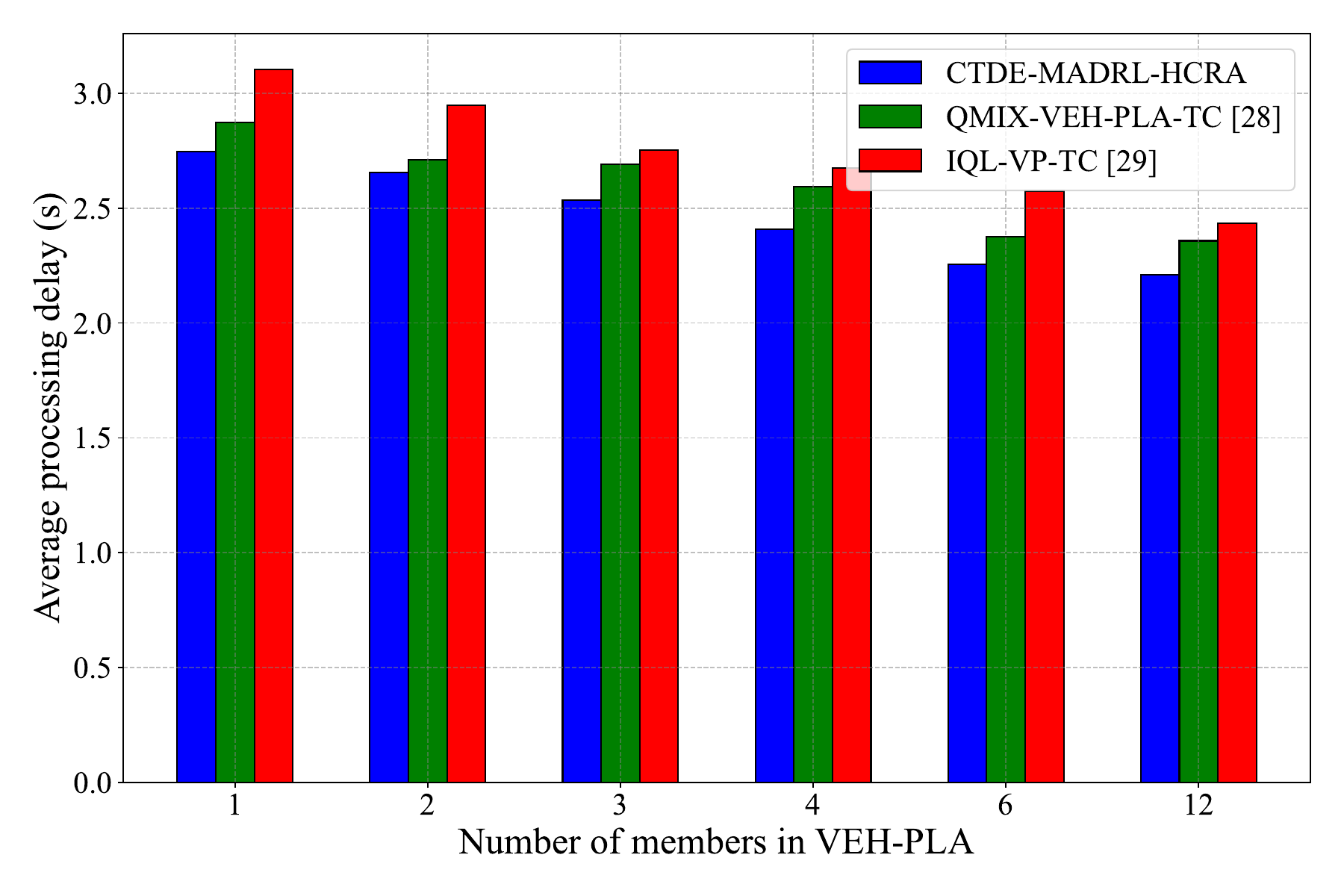} % 图片文件的路径
  \caption{Members in VEH-PLA vs. average processing delay} % 图片的标题
  \label{fig:相同} % 用于引用图片的标签
\end{figure}

%For the better observation, the results are averaged with the number of episodes varying from 500 to 1000. 
The number of moving vehicles on the road is set to 15, and other parameters are set as before. I particular, considering 12 VEH-PLA members, which are classified into 1, 2, 3, 4, and 6 platooning respectively, and 3 non-platooning members, Fig. \ref{fig:相同} describes the performance comparison of average processing delay under different number of platooning, The results clearly indicate that \textbf{CTDE-MADRL-HCRA} is more applicable for various platoon modes, followed by \textbf{QMIX-VEH-PLA-TC}, the corresponding reason is similar as that of Fig. \ref{fig:Accumulated Average Reward}, Fig. \ref{fig:latency}, and Fig. \ref{fig:车数}. Similarly, \textbf{IQL-VEH-PLA-TC} exhibits the worst performance. Given a certain number of moving vehicles, decreasing the number of VEH-PLA (i.e., increasing the number of vehicles in a single VEH-PLA) can have actually helped reduce the average processing delay. This is attributed to the fact that more fewer VEH-PLA can lead to more fewer number of public task offloading. However, the size of VEH-PLA is not more larger the better, one possible reason is that increased communication overhead and coordination difficulties among a larger number of vehicles can lead to inefficiencies and delays in the platooning network, and has been concluded in \cite{9878288}, \cite{10084415}. 
Therefore, in this case, keeping the size of VEH-PLA with the number of 6 members can make a better trade-off between the average processing delay and platooning stability.

\begin{figure}[htbp]
  \centering % 让图片居中显示
  \includegraphics[width=0.32\textwidth]{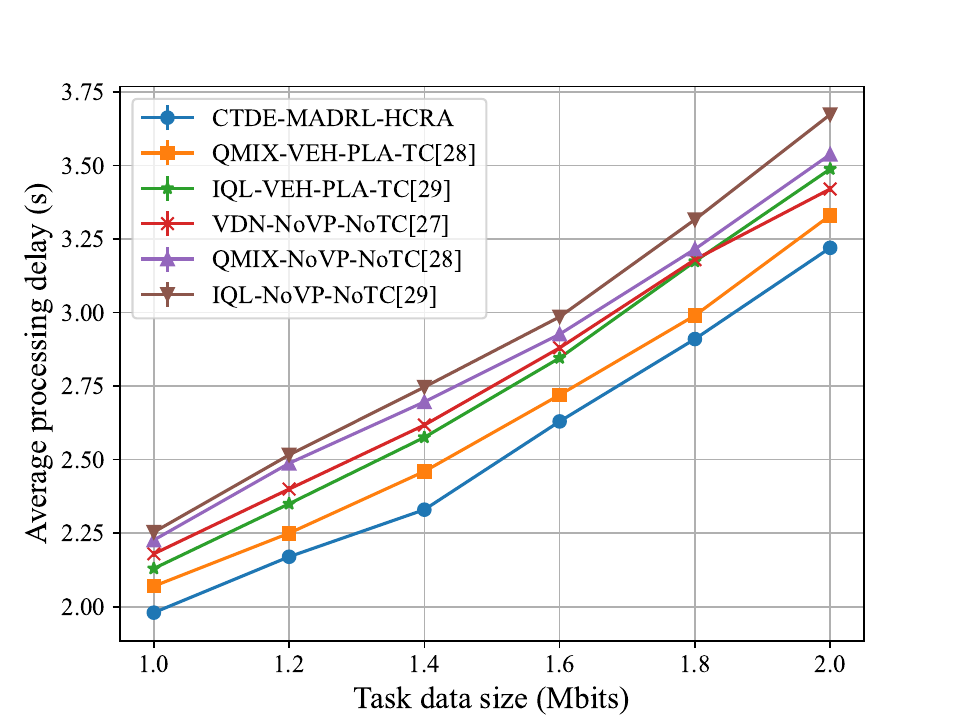} % 图片文件的路径
  \caption{Size of task data vs. average processing delay  } % 图片的标题
  \label{fig:Task data size} % 用于引用图片的标签
\end{figure}

\iffalse
\begin{table}[ht]
\centering
\caption{Growth rate of average delay over task size}
\label{tab:slopes}
\begin{tabularx}{\linewidth}{@{}Xccccc@{}}
\toprule
Schemes & 1.0-1.4 & 1.4-1.6 & 1.6-1.8 & 1.8-2.0 & Avg. \\
\midrule
CTDE-MADRL-HCRA & 0.87 & 1.50 & 1.40 & 1.55 & 1.24 \\
QMIX-VEH-PLA-TC \textsuperscript{\cite{rashid2020monotonic}} & 0.97 & 1.30 & 1.35 & 1.70 & 1.26 \\
IQL-VEH-PLA-TC \textsuperscript{\cite{tampuu2017multiagent}} & 1.11 & 1.35 & 1.64 & 1.57 & 1.36 \\
VDN-NoVP-NoTC \textsuperscript{\cite{10147826}} & 1.10 & 1.31 & 1.50 & 1.20 & 1.24 \\
QMIX-NoVP-NoTC \textsuperscript{\cite{rashid2020monotonic}} & 1.17 & 1.15 & 1.45 & 1.61 & 1.31 \\
IQL-NoVP-NoTC \textsuperscript{\cite{tampuu2017multiagent}} & 1.23 & 1.19 & 1.65 & 1.79 & 1.42 \\
\bottomrule
\end{tabularx}
\end{table}
\fi

%For the better observation, the results are averaged with the number of episodes varying from 500 to 1000. 
The size of of task data varies in range of [1.0, 2.0]Mbits, and other parameters are set as before. Fig. \ref{fig:Task data size} validates the impact of task size on the average processing delay.  \textbf{CTDE-MADRL-HCRA} shows the best performance as expected. In particular, given a task data with the size of 2 Mbits, compared with \textbf{QMIX-VEH-PLATC} and \textbf{IQL-VEH-PLA-TC}, the average processing delay achieved by \textbf{CTDE-MADRL-HCRA} is decreased 110 ms and 240 ms, respectively. %Furthermore, it can be seen from Table ~\ref{tab:slopes} that the average slopes of offloading schemes based on IQL (i.e., \textbf{IQL-VEH-PLA-TC} and \textbf{IQL-NoVP-NoTC}) are relatively higher, the corresponding reason is similar to these of Fig. \ref{fig:Accumulated Average Reward}, Fig. \ref{fig:latency}, and Fig. \ref{fig:车数}.

\section{CONCLUSION AND FUTURE WORK}
\label{sec:CONCLUSION AND FUTURE WORK}
In this paper, by integrating the resources of individual vehicle with the concept of VEH-PLA, along with the resources of RSUs and MBSs, we investigated the task offloading in the IoV characterized by the mobility and wireless interference, with the purpose of minimizing the processing delay and keeping the resource load balancing. A DRL framework based on CTDE was introduced to enhance the accuracy of task offloading and resource utilization. The effectiveness of proposed offloading scheme was further demonstrated by simulation results, which made a comprehensive comparison with other baseline offloading schemes by considering the influences of the number of episodes and vehicles, the size of VEH-PLA and task data, respectively. In particular, the proposed scheme showed the lowest processing delay. 

Owing to space limitations, this paper aimed to minimize the processing delay of task offloading and keep the load balancing of resources. In fact, it can be noticed that the performance of task offloading depends on the vehicular mobility and energy consumption. 
Therefore, in our future work, we will
devote into joint optimization of the resource allocation and vehicle trajectory prediction to minimize the processing delay and energy consumption simultaneously. In addition, the integration of
unmanned aerial vehicles (UAV) into the IoV can further enhance
the network coverage and adjust computational resource
allocation more flexibly, which also deserves further investigation.

\iffalse
In this work, we address the contradiction between high data generation rates and limited onboard computing resources in ICVs by proposing a task offloading mechanism within a heterogeneous edge computing network based on VEH-PLA. Utilizing a DRL framework grounded in CTDE, this mechanism effectively enhances resource optimization and QoS improvement. Through task categorization and the introduction of the VEH-PLA concept, not only have we reduced the frequency of task offloading, but we've also alleviated resource scarcity issues through intelligent scheduling strategies, thus elevating overall system efficiency. The effectiveness of our proposed method is corroborated by simulation results, which demonstrate significant improvements in system efficiency and resource management compared to traditional methods and other baseline algorithms.Despite the achievements of this research, the evolution of intelligent vehicular networks presents many challenges that warrant further exploration. This paper specifically optimizes the critical delay objective within task offloading issues. Future work could expand into multi-objective optimization concerning energy consumption, data security, etc., and investigate task offloading and resource sharing mechanisms across different network scenarios (e.g., urban, highways) to enhance the availability and reliability of vehicular network services.
\fi

\bibliography{bibtex/bib/IEEEabrv,bibtex/bib/references}

\end{document}